\documentclass[twocolumn]{aastex631}

\usepackage{amssymb}
\usepackage{mathrsfs}
\usepackage{amsmath}

\begin{document}

\title{Revealing an Oscillating and Contracting Compact Corona near the Event Horizon of the Supermassive Black Hole in 1ES 1927+654}

\correspondingauthor{Qing-Cang Shui}
\email{shuiqc@ihep.ac.cn}
\correspondingauthor{Shu Zhang}
\email{szhang@ihep.ac.cn}
% \correspondingauthor{Hua Feng}
% \email{hfeng@ihep.ac.cn}

\author[0000-0001-5160-3344]{Qing-Cang Shui}
\affiliation{Key Laboratory of Particle Astrophysics, Institute of High Energy Physics, Chinese Academy of Sciences, 100049, Beijing, China}
\affiliation{University of Chinese Academy of Sciences, Chinese Academy of Sciences, 100049, Beijing, China}

% \author{Your Name}
% \affiliation{Key Laboratory of Particle Astrophysics, Institute of High Energy Physics, Chinese Academy of Sciences, 100049, Beijing, China}
% \affiliation{University of Chinese Academy of Sciences, Chinese Academy of Sciences, 100049, Beijing, China}
% \affiliation{Institut f\"{u}r Astronomie und Astrophysik, Kepler Center for Astro and Particle Physics, Eberhard Karls, Universit\"{a}t, Sand 1, D-72076 T\"{u}bingen, Germany}
\author{Shu Zhang}
\affiliation{Key Laboratory of Particle Astrophysics, Institute of High Energy Physics, Chinese Academy of Sciences, 100049, Beijing, China}

\author[0000-0001-5586-1017]{Shuang-Nan Zhang}
\affiliation{Key Laboratory of Particle Astrophysics, Institute of High Energy Physics, Chinese Academy of Sciences, 100049, Beijing, China}
\affiliation{University of Chinese Academy of Sciences, Chinese Academy of Sciences, 100049, Beijing, China}

\author[0000-0001-7584-6236]{Hua Feng}
\affiliation{Key Laboratory of Particle Astrophysics, Institute of High Energy Physics, Chinese Academy of Sciences, 100049, Beijing, China}

\author[0000-0001-8768-3294]{Yu-Peng Chen}
\affiliation{Key Laboratory of Particle Astrophysics, Institute of High Energy Physics, Chinese Academy of Sciences, 100049, Beijing, China}

\author[0000-0001-9599-7285]{Long Ji}
\affiliation{School of Physics and Astronomy, Sun Yat-Sen University, Zhuhai, 519082, China}

\author[0000-0003-3188-9079]{Ling-Da Kong}
\affiliation{Institut f\"{u}r Astronomie und Astrophysik, Kepler Center for Astro and Particle Physics, Eberhard Karls, Universit\"{a}t, Sand 1, D-72076 T\"{u}bingen, Germany}

\author[0000-0003-4498-9925]{Liang Zhang}
\affiliation{Key Laboratory of Particle Astrophysics, Institute of High Energy Physics, Chinese Academy of Sciences, 100049, Beijing, China}

\author[0000-0002-5554-1088]{Jing-Qiang Peng}
\affiliation{Key Laboratory of Particle Astrophysics, Institute of High Energy Physics, Chinese Academy of Sciences, 100049, Beijing, China}

\author[0000-0002-6454-9540]{Peng-Ju Wang}
\affiliation{Institut f\"{u}r Astronomie und Astrophysik, Kepler Center for Astro and Particle Physics, Eberhard Karls, Universit\"{a}t, Sand 1, D-72076 T\"{u}bingen, Germany}

%\affiliation{Yunnan Observatories, Chinese Academy of Sciences, Kunming 650216, P.R. China}


\begin{abstract}
Dynamic processes in the accretion flow near black holes produce X-ray flux variability, sometimes quasi-periodic. Determining its physical origin is key to mapping accretion geometry but remains unresolved. We perform a novel phase-resolved analysis on a newly discovered quasi-periodic oscillation (QPO) in the active galactic nucleus 1ES 1927+654. For the first time in a supermassive black hole (SMBH), we detect a unique `U'-shaped QPO lag-energy spectrum and observe coronal spectral variability over the QPO phase. We find that the QPO is adequately explained by plasma resonant oscillations within a corona. Modeling of QPO spectral properties and reverberation mapping reveal that the corona is contracting and confined to only a few gravitational radii regions near the SMBH, consistent with theoretical predictions for a decreasing QPO period of near 10 minutes. These results present the first observational evidence for an oscillating and contracting compact corona around an SMBH.
\end{abstract}



\section{Introduction} \label{sec:intro}
Black holes powered by extreme accretion of material are associated with some of the most luminous phenomena in the universe---like active galactic nuclei (AGNs), tidal disruption events (TDEs), and gamma-ray bursts \citep{1964ApJ...140..796S,1988Natur.333..523R,1999ApJ...518..356P}. In AGNs, X-rays are emitted from the innermost regions of the accretion flow which govern the formation of hot coronae and launching of powerful plasma outflows \citep{2019ARA&A..57..467B}. It is still unclear why the X-ray flux usually exhibits extreme variability \citep{2025ARA&A..63..379K}, which is sometimes quasi-periodic. Notably, high-frequency ($\sim 2000[M_{\bullet}/M_{\odot}]^{-1}$ Hz, where $M_{\bullet}$ is the black hole mass, and $M_{\odot}$ is the mass of the sun) X-ray quasi-periodic oscillations (QPOs) have been revealed in accreting black holes spanning seven orders of magnitude in mass \citep{2015ApJ...798L...5Z}. X-ray variability has been proposed as a valuable proxy for indirectly mapping structures of the innermost accretion flows \citep{2014A&ARv..22...72U,2016MNRAS.462..511K,2019Natur.565..198K}, which are compact regions difficult to resolve with current X-ray telescopes. However, to fully harness this potential, it is essential to determine the physical origin of the observed variability, which remains a big puzzle. For high-frequency QPOs (HFQPOs), while typical timescales in black hole X-ray binaries (BHXBs) are too short to be well investigated with current X-ray telescopes \citep{2013MNRAS.435.2132M}, longer QPO time scales are expected in supermassive black holes (SMBHs) in AGNs. Thus, the recent discovery of a $\sim$10-minute X-ray QPO from AGN 1ES 1927+654 \citep{2025Natur.638..370M} provides a unique laboratory to probe both the origin of the QPO and dynamic processes in the accretion flow.

The nearby ($z=0.019422$) AGN 1ES 1927+654 hosts an SMBH with mass $M_{\bullet}=1.38^{+1.25}_{-0.66}\times10^6\ M_{\odot}$ \citep{2022ApJ...933...70L}. Originally classified as a typical ``naked" type 2 AGN due to its persistent absence of broad emission lines \citep{2003A&A...397..557B}, this source underwent a major outburst detected by \textit{All-Sky Automated Survey for Supernovae} (\textit{ASAS-SN}) in March 2018, exhibiting a V-band flux increase exceeding two magnitudes. Over the following three years, optical spectroscopic observations revealed the emergence of broad Balmer lines within a few weeks of the outburst, followed by the delayed appearance of broad Ly$\alpha$ emission on 2018 August 28, making this the first AGN in which a transition between different spectral types was temporally resolved \citep{2019ApJ...883...94T}. Following the outburst, ultraviolet (UV) photometry revealed a characteristic $t^{-0.91\pm0.04}$ power-law delay, recovering to near pre-outburst levels after $\sim1200$ days \citep{2022ApJ...931....5L}. 

The X-ray band, however, displayed non-monotonic evolution (see Fig. \ref{fig:Figure1}A). About 3--4 months after the optical flare, X-rays began to dim, and by 2018 August, the 2--10 keV component had entirely vanished \citep{2020ApJ...898L...1R}. A few months later, the X-ray flux reemerged at up to ten times its pre-outburst level, then underwent a year-long gradual decline accompanied by a strengthening of the X-ray corona \citep{2022ApJ...934...35M}. Radio observations prior to 2023 showed that 1ES 1927+654 displayed properties typical of radio-quiet Seyfert galaxies \citep{2022ApJ...931....5L}. In February 2023, however, the source began exhibiting a radio flare with a steep exponential rise, ultimately reaching a flux level $\sim$60 times higher than before, and it has maintained this enhanced emission for over a year \citep{2025ApJ...979L...2M,2025ApJ...981..125L}. \textit{Very Long Baseline Array} (\textit{VLBA}) observations revealed spatially resolved, bipolar jet-like extensions on scales of 0.1--0.3 pc, indicating a new, mildly relativistic outflow \citep{2025ApJ...979L...2M}. 

In early 2022, approximately 200 days before the onset of the radio flare, a distinctive X-ray rebrightening emerged without corresponding optical/UV counterparts \citep{2023ApJ...955....3G}. During this increase, \textit{XMM-Newton} observations from July 2022 to October 2024 detected a millisecond (mHz) QPO \citep{2025Natur.638..370M}, temporally coincident with the jet formation (see Fig.~\ref{fig:Figure1}A). This QPO exhibits unique frequency evolution (see Fig.~\ref{fig:Figure1}B), contrasting with stable-frequency QPOs observed in other SMBHs, such as the super-Eddington narrow-line Seyfert 1 AGN RE J1034+396 \citep{2008Natur.455..369G}, the canonical tidal disruption event (TDE) ASASSN-14li \citep{2019Sci...363..531P}, and the TDE candidate 2XMM J1231 \citep{2013ApJ...776L..10L}.

In this Letter, we perform a novel phase-resolved analysis on the newly discovered unique QPO in 1ES 1927+654 using \textit{XMM-Newton} observations. We provide an overview of the observations and our data reduction in Section~\ref{sec:2}, followed by the presentation of our results in Section~\ref{sec:3}. Finally, we discuss and summarize these results in Sections~\ref{sec:4} and \ref{sec:5}, respectively.

\section{Observations and Data Reduction} \label{sec:2}
The \emph{XMM-Newton} observatory has extensively monitored 1ES 1927+654 since 2011, including a pre-outburst observation in 2011 and nineteen observations following the onset of the 2018 outburst. In this study, we focused specifically on observations in which the mHz QPO was detected. Eight of these observations have been previously reported \citep{2025Natur.638..370M}, while an additional four are newly presented in this work (see Table~\ref{tab:S1}).

\begin{table*} % Do not use \begin{table*}
	\centering
	% Captions go above tables
	\caption{Details of the QPO in 2--10 keV \textit{XMM-Newton} data. ${\rm ^a}$Centroid frequency of the QPO, as measured from the PDS fitting analysis. ${\rm ^b}$Quality factor of the QPO, defined as $Q=f_{\rm QPO}/{\rm FWHM_{QPO}}$, as measured from the PDS fitting analysis. ${\rm ^c}$Fractional RMS of the QPO, as measured from the normalization of the additional Lorentzian for the QPO. ${\rm ^d}$Parameter values directly taken from the public work \citep{2025Natur.638..370M}. Uncertainties are given at the 68 per cent ($1\sigma$) confidence level.}
	\label{tab:S1} % give each table a logical label name
	\begin{tabular}{ccccc} % four columns, alignment for each
		\\
\hline
Epoch & obsID (s) & ${f_{\rm QPO}}^{\rm a}$ (mHz) & ${Q_{\rm QPO}}^{\rm b}$ & $\rm {RMS_{QPO}}^{c}$ (\%)\\
\hline
Jul.-Aug. 2022 & 0902590201-501${\rm ^d}$ & $0.93\pm0.06$ & $2.3^{+1.0}_{-0.5}$ & $12.1^{+1.7}_{-1.5}$\\
Feb. 2023 & 0915390701${\rm ^d}$ & $1.67\pm0.04$ & $8.7^{+5.4}_{-3.0}$ & $19.1^{+3.4}_{-2.5}$\\
Aug. 2023 & 0931791401${\rm ^d}$ & $2.21\pm0.05$ & $8.0^{+4.5}_{-2.6}$ & $15.2^{+2.3}_{-1.9}$\\
Mar. 2024 & 0932392001${\rm ^d}$ & $2.34\pm0.05$ & $10.1^{+13.8}_{-4.7}$ & $15.6^{+4.1}_{-2.9}$\\
Mar. 2024 & 0932392101${\rm ^d}$ & $2.50\pm0.18$ & $7.8^{+12.2}_{-2.8}$ & $7.6^{+2.3}_{-2.6}$\\
Jul. 2024 & 0953010401 & $2.41^{+0.10}_{-0.06}$ & $8.0^{+12.9}_{-3.6}$  & $11.9^{+2.7}_{-2.3}$\\
Jul. 2024 & 0953010501 & $2.42^{+0.14}_{-0.12}$ & $4.3^{+3.4}_{-1.3}$ & $9.5\pm2.0$\\
Oct. 2024 & 0953010901 & $2.50^{+0.26}_{-0.25}$ & $4.6^{+9.2}_{-1.9}$ & $13.0^{+4.0}_{-3.8}$\\
Oct. 2024 & 0953010601 & $2.42\pm0.04$ & $16.0^{+25.7}_{-8.7}$ & $13.4^{+3.7}_{-2.6}$\\

 \hline
	\end{tabular}
\end{table*}

Data reduction was performed using the \emph{XMM-Newton} Science Analysis System (\textsc{xmmsas}) version 19.1.0 with the latest calibration files (February 2021). Raw datasets were retrieved from the \emph{XMM-Newton} Science Archive (XSA). For this analysis, we utilized data exclusively from the European Photon Imaging Camera (EPIC) pn detector due to its superior sensitivity. The standard \textsc{epproc} pipeline was employed to generate EPIC-pn event files. For timing analysis, source photons were extracted from a circular region of radius $35''$ centered on the source position, with events selected under the criterion \texttt{PATTERN $\leq 4$} for the EPIC-PN data. For spectral analysis, where extremely high count rates were not required, we used an annular extraction region with inner and outer radii of $10''$ and $40''$, respectively, to mitigate mild pile-up effects \citep{2025Natur.638..370M}. The background was estimated using a nearby region with radius of $35''$ free from the source position. To identify high-background periods, we generated light curves in the 10–15 keV band. No significant background flares were detected in the continuous data segments. Background-subtracted light curves for 1ES 1927+654 were produced using the \textsc{epiclccorr} task. We verified that background contributions did not affect the corrected light curves, with the source-to-background flux ratio consistently exceeding 15 within the source extraction region.

For spectral analysis, response matrices and ancillary response files were created using the \textsc{xmmsas} tasks \textsc{rmfgen} and \textsc{arfgen}. Finally, the source spectra were grouped using the tool \textsc{specgroup} to a minimum of 25 counts per spectral bin and at the same time not to oversample the instrument energy resolution by more than a factor of 3. This allows the application of $\chi^2$ statistics in the spectral modeling. 

The Neutron star Interior Composition Explorer (\emph{NICER}) has been monitoring 1ES 1927+654 with a variable cadence, ranging from two observations per day to one or two observations per week, since shortly after the onset of the 2018 outburst. Fig.~\ref{fig:Figure1}A presents the \emph{NICER} light curves in two energy bands. These light curves were directly obtained from the public work \citep{2025Natur.638..370M}, which has made the datasets publicly available\footnote{ \url{https://github.com/memasterson/1ES1927_mHzQPO.git}}.

\section{Analysis and Results}
\label{sec:3}
\begin{figure*}
\centering
    \includegraphics[width=\linewidth]{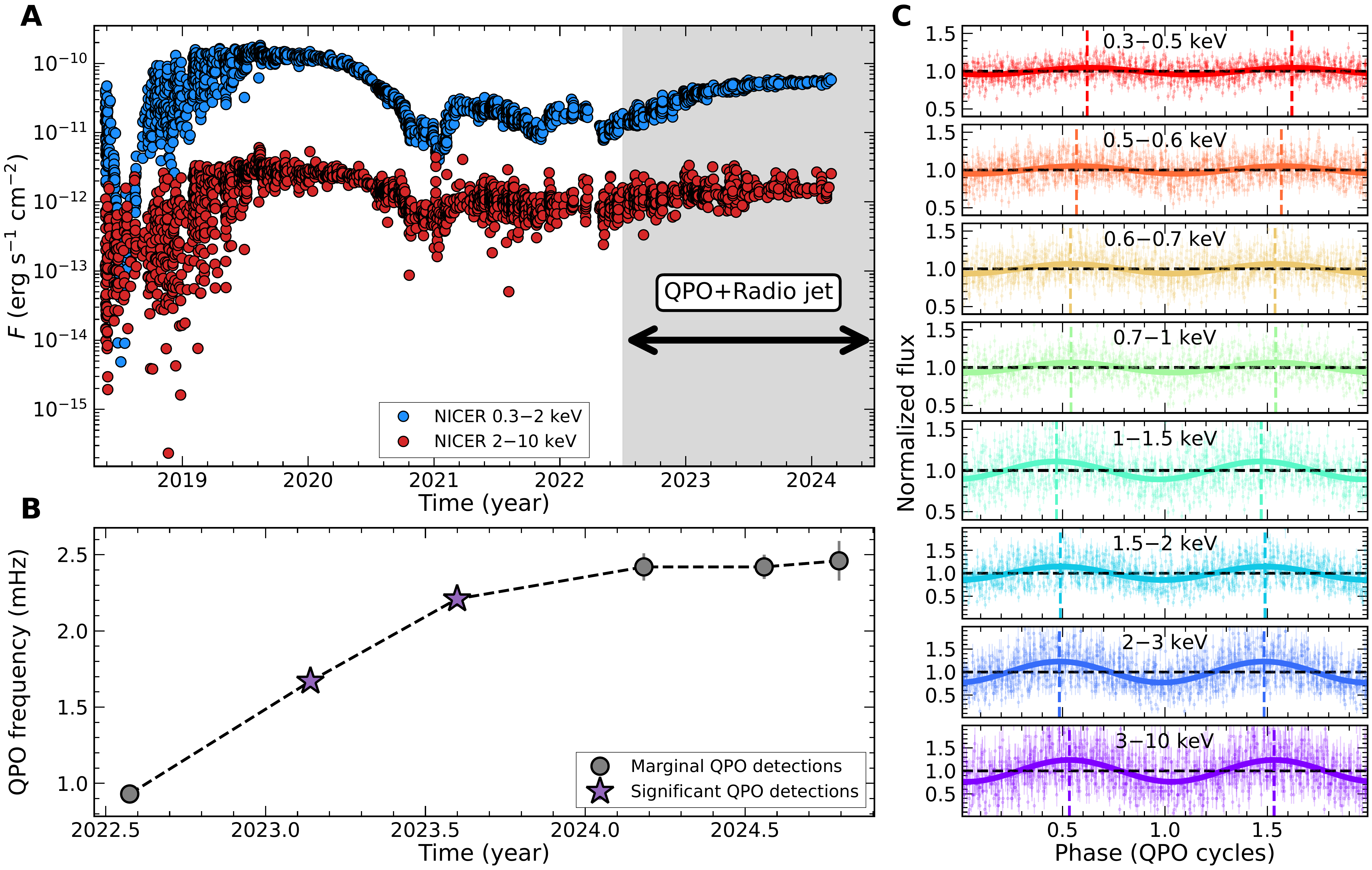}
    \caption{Overview of the X-ray QPO detected in 1ES 1927+654. (\textbf{A}) Long-term \emph{NICER} light curves in energy ranges of 0.3--2 keV (blue circles) and 2--10 keV (red circles) beginning 2 months after the optical outburst \citep{2025Natur.638..370M}. The epoch of the QPO detected by \emph{XMM-Newton} is shaded in gray. The timing of the mHz QPO detections suggests a possible association with the formation of the jet, as inferred from radio observations \citep{2025ApJ...979L...2M,2025ApJ...981..125L}. (\textbf{B}) Evolution of the QPO frequency over time. Purple stars indicate significant detections of the QPO ($>6\sigma$), while gray circles denote marginal detections ($<6\sigma$). The presented QPO frequency and corresponding $1\sigma$ error bars are obtained from the PSD fitting with an additional Lorentzian. (\textbf{C}) Normalized phase-folded light curves (QPO waveforms) of the $\sim1.67$ mHz QPO (obsID 0915390701) across multiple energy bands, referenced to the 2--10 keV QPO phase. Data points with error bars are unbinned waveforms, while solid lines are best-fit sinusoidal functions. Data are plotted over two QPO cycles by repeating the pattern. Vertical dashed lines indicate the peak phases of the QPO waveforms obtained from sinusoidal fittings.}
    \label{fig:Figure1}
\end{figure*}

We first conducted a standard power spectral density (PSD) analysis with \textit{XMM-Newton} data following the procedure from the published work \citep[see Appendix~\ref{appendix1} and][for details]{2025Natur.638..370M}. Table \ref{tab:S1} presents detailed QPO parameters obtained from the PSD modeling.
Specifically, the QPO frequency significantly increases from July 2022 to August 2023 before reaching a plateau approximately one year after its initial detection (see Fig.~\ref{fig:Figure1}B). Previous PSD modeling suggested a tentative energy-dependent increase in QPO amplitude \citep{2025Natur.638..370M}, but the physical mechanism remained unclear due to limitations of conventional PSD analysis. To glean insights beyond the limits of conventional PSD analysis, we employed a novel technique to perform phase-resolved analysis (see Appendix~\ref{appendix2} and \ref{appendix3} for details) on two \textit{XMM-Newton} observations (February and August 2023, corresponding observational IDs are 0915390701 and 0931791401) that respectively show QPO detections at frequencies of $\sim$1.67 and $\sim$2.21 mHz with $\gtrsim6\sigma$ significance \citep{2025Natur.638..370M}. Folding the light curves with QPO phase reveals that the flux modulation is significant in all used bands in the 0.3--10 keV energy range, with clear energy dependence (see Fig.~\ref{fig:Figure1}C). In particular, the QPO amplitude increases with energy, and peaks of high-energy ($\gtrsim1$ keV) waveforms precede those of low-energy ($\lesssim1$ keV) waveforms, indicating QPO time lags across energy bands.

\begin{figure*}
\centering
\includegraphics[width=0.75\linewidth]{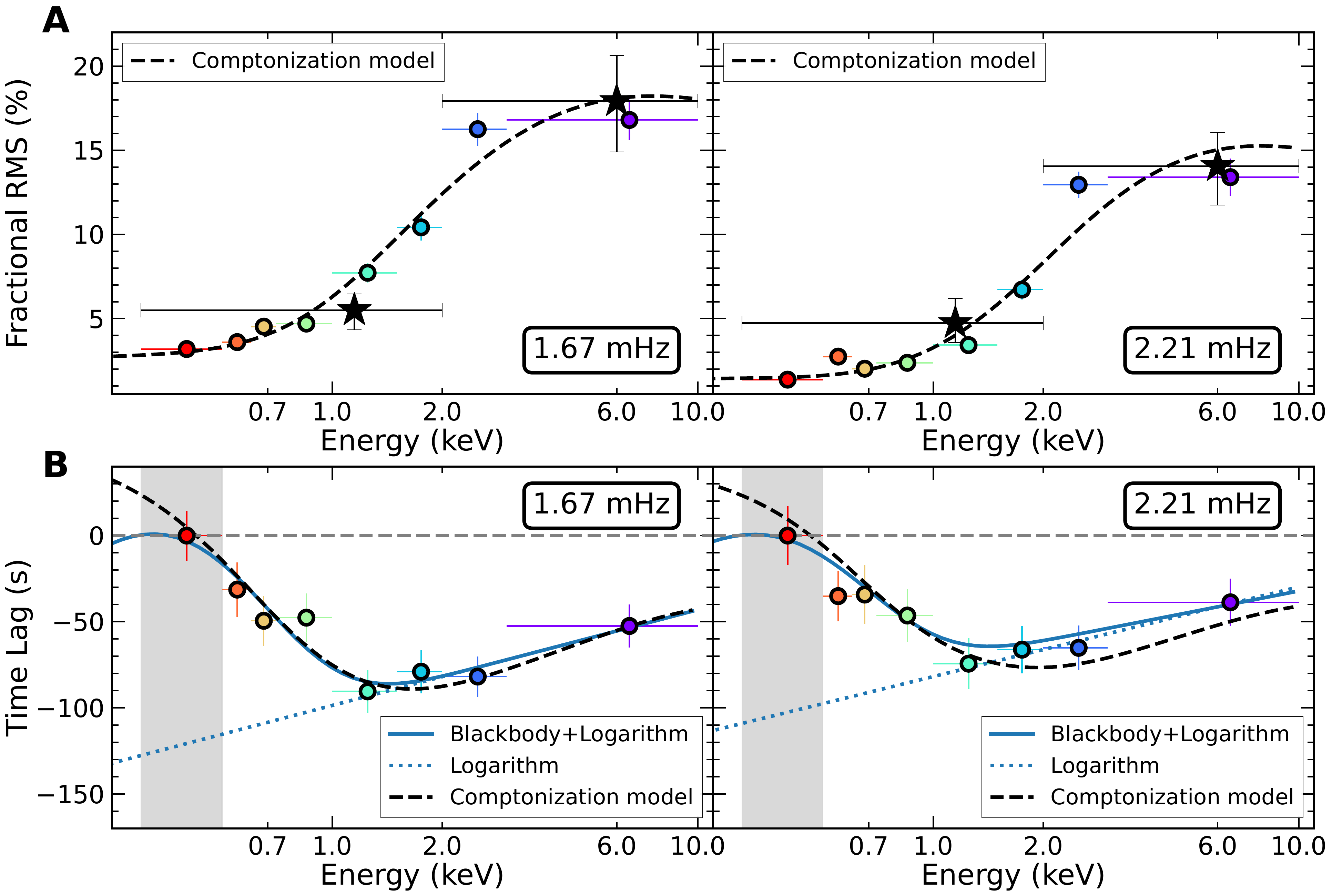}
    \caption{RMS- and lag-energy spectra of the X-ray QPO in 1ES 1927+654 and corresponding fitting models. Error bars represent $1\sigma$ confidence intervals. The black dashed lines show the best-fit results from the joint fitting of time-averaged spectra, QPO RMS- and lag-energy spectra with the time-dependent Comptonization model. (\textbf{A}) QPO RMS-energy spectra. Black stars represent values from conventional PSD modeling using a Lorentzian function; colored circles show results from the phase-resolved analysis. (\textbf{B}) QPO lag-energy spectra with the 0.3--0.5 keV energy band as the reference (gray shaded region), defined as zero lag. The uncertainty in the reference band is based on the arithmetic mean of the $1\sigma$ errors in other energy bands. The blue solid lines represent a model with a soft-blackbody component and a hard-logarithmic power-law component, and the blue dotted lines show the corresponding hard-logarithmic component. \label{fig:Figure2}}
\end{figure*} 

The high-significance QPO waveforms obtained from our phase-folded analysis enable precise determination of the energy-dependent QPO fractional root mean square (RMS) amplitude and time lag (see Appendix~\ref{appendix3}), surpassing capabilities of the traditional Fourier transform analysis. Figure \ref{fig:Figure2}A shows RMS-energy spectra for the two observations, comparing results from the phase-folded method with those from the traditional PSD modeling analysis. While both methods show consistent trends, our phase-folded analysis resolves finer spectral details: the QPO RMS amplitude slightly increases with energy below 1 keV before sharply increasing by a factor of two (from $\sim$5\% to $\sim$15\%) from $\sim$1 to $\sim$3 keV. Figure \ref{fig:Figure2}B presents the lag-energy spectra, revealing, for the first time among both AGNs and BHXBs, a clear `U'-shaped profile for an HFQPO. We observe a significant thermal soft lag below 1 keV, followed by a hard lag at higher energies. These two distinct lag-energy behaviors are attributed to different physical mechanisms: the thermal soft lag arises from the reprocessing of Comptonization photons in an irradiated cold disk \citep{2014A&ARv..22...72U,2017MNRAS.471.1475D}, while the hard lag reflects longer photon Compton upscattering timescales at higher energies \citep{1988Natur.336..450M,1997ApJ...480..735K,2022MNRAS.515.2099B}. In this case, hard lags would exhibit a logarithmic dependency on photon energy $E$, with the time lag between energies $E_1$ and $E_2$, $\Delta\tau\propto\ln{(E_2/E_1)}$ \citep{1999ApJ...510..874N}. The lag-energy spectra are well-described by a model that combines the soft-blackbody and hard-logarithmic components (Fig.~\ref{fig:Figure2}B). The best-fitting parameters are listed in Table \ref{tab:1}. For consistency with previous time-lag studies \citep{2017MNRAS.471.1475D,2019Natur.565..198K,2020ApJ...899...44W}, we measured the amplitude of the thermal reverberation lag as the difference between the hard component and the maximum lag below 1 keV. The thermal reverberation lags found in this way are $121\pm14$ s and $101\pm17$ s, or $(10.4\pm1.2)R_{\rm g}/c$ and $(8.7\pm1.5)R_{\rm g}/c$ (where $R_{\rm g}=GM_\bullet/c^2$, assuming the black hole mass, $M_\bullet=1.38 \times 10^6 M_{\odot}$ and the disk inclination, $i=45^\circ$) for the 1.67-mHz and 2.21-mHz QPO observations, respectively, showing a slight decrease with the increase of QPO frequency. Additionally, the observed thermal soft lags and the measured black hole mass of 1ES 1927+654 \citep{2022ApJ...933...70L} are consistent with the correlation between the two parameters observed in multiple radio-quiet AGNs \citep{2013MNRAS.431.2441D} (see Fig. \ref{fig:S5}).

Notably, both the RMS- and lag-energy spectra closely resemble those of low-frequency QPOs (LFQPOs) observed in the intermediate states of BHXBs \citep{2020MNRAS.496.4366B,2023MNRAS.525..854M}, suggesting that radiative properties of this X-ray mHz QPO are similarly dominated by Comptonization processes in a hot corona. For the first time, thanks to the precise determination of the QPO RMS- and lag-energy spectra from our phase-folded analysis, we applied the time-dependent Comptonization model \citep{2022MNRAS.515.2099B}, a QPO radiative model widely used in BHXB LFQPO studies \citep[e.g.][]{2020MNRAS.496.4366B,2023MNRAS.519.1336P,2023MNRAS.525..854M}, to analyze an HFQPO in an AGN (see Appendix~\ref{appendix4} for details). In this model, seed photons from the accretion disk, with an inner temperature of $kT_{\rm s}$, undergo Comptonization in a spherical corona with the size $L$ and electron temperature $kT_{\rm e}$. The model also incorporates feedback processes, where a fraction of the Comptonized photons are redirected back onto the accretion disk and reprocessed. The reprocessing fraction $\eta$ represents the fraction of the disk flux resulting from coronal irradiation. This model does not depend on a specific dynamic origin of the QPO (e.g. QPO frequency), but instead treats it as small oscillation of the spectrum around the time-averaged one. We observe that both of the RMS- and lag-energy spectra are well-described by the time-dependent Comptonization model (see Fig.~\ref{fig:Figure2}). By jointly fitting the time-averaged spectra, QPO RMS- and lag-energy spectra with the model (see Fig.~\ref{fig:S7}), we estimated the corona size to be $7.1^{+1.1}_{-0.8}R{\rm g}$ and $4.0^{+2.1}_{-1.8}R{\rm g}$ for the two observations, respectively. These estimates are consistent with results obtained from modeling the thermal reverberation lag, indicating a highly compact contracting corona, close to the event horizon. We notice that so far the similar studies of LFQPOs and thermal reverberation lags in most Galactic BHXBs suggested coronae much larger ($>100R_{\rm g}$) than those inferred from reflection modeling \citep{2017MNRAS.471.1475D,2020ApJ...899...44W,2022MNRAS.515.2099B}. Details on the fitting parameters for the time-dependent Comptonization model can be found in Table~\ref{tab:1}. 

\begin{figure*}
\centering
    \includegraphics[width=0.75\linewidth]{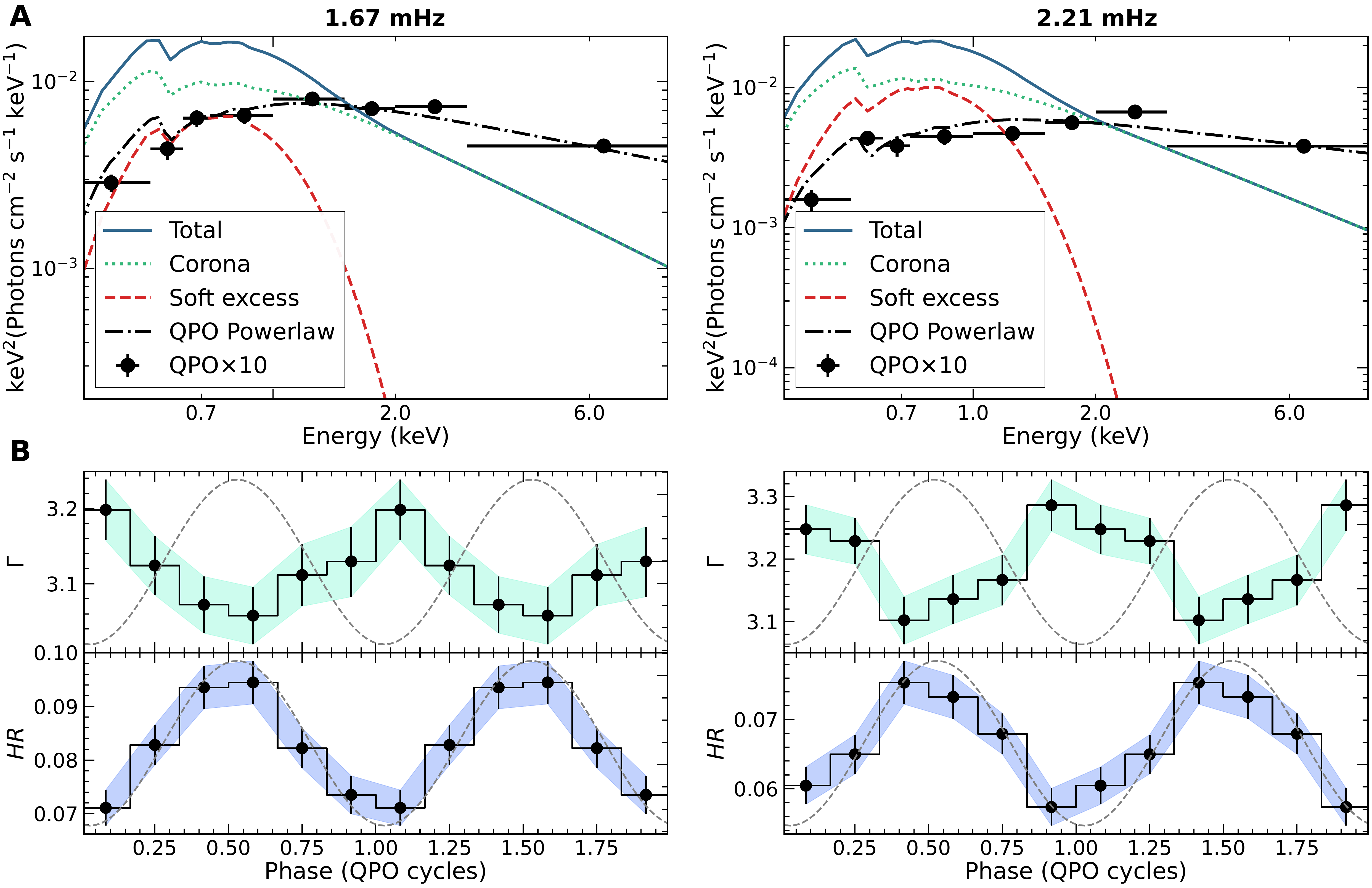}
    \caption{Spectral properties of the QPO in 1ES 1927+654. (\textbf{A}) Comparison of QPO spectra with time-averaged energy spectra. Individual model components are indicated by different line styles and colors. QPO spectra and their corresponding fit models are scaled by a factor of 10 for better comparison. (\textbf{B}) Variability of the spectral index and hardness ratio over the QPO cycle. The hardness ratio is defined as the flux ratio between the 2--10 keV and 0.3--2 keV energy bands. The phase-folded QPO light curve in 2--10 keV is plotted in each panel as a gray dashed line for reference. \label{fig:Figure3}}
\end{figure*}

We constructed QPO spectra by converting the QPO RMS-energy spectra into flux-equivalent units through the instrument response calibration. It is evident that QPO spectra are significantly harder than both the time-averaged total and Comptonization spectra (Fig.~\ref{fig:Figure3}A). Spectral fittings of the QPO spectra with a single power-law model, using the same interstellar absorption as the time-averaged spectra, yield spectral indices of $2.43 \pm 0.06$ and $2.42 \pm 0.08$ for the two observations, respectively, which are significantly smaller than those of the time-averaged Comptonization spectra ($3.09 \pm 0.05$ and $3.19 \pm 0.05$). Monte Carlo radiative transfer simulations using the code \textsc{monk} \citep{2019ApJ...875..148Z} demonstrate that the observed QPO spectral hardening, compared to the time-averaged spectra, arises from oscillations in the coronal temperature and/or optical depth, rather than oscillations in the accretion rate of the disk (see Fig. \ref{fig:S8} in Appendix~\ref{appendix5} for details). These coronal oscillations are further supported by phase-resolved spectral analysis (see Appendix~\ref{appendix3}), which reveals significant energy spectral variability over the QPO cycle. Specifically, the coronal spectrum hardens during the QPO peaks, as indicated by an anti-correlation between spectral index and flux, and by the synchronous modulation of hardness ratio with flux (Fig.~\ref{fig:Figure3}B).

\begin{table} % Do not use \begin{table*}
	\centering
	% Captions go above tables
	\caption{Fit parameters of models used to model the RMS- and lag-energy spectra of the QPO in 1ES 1927+654. Observations 1 and 2 represent QPO detections at frequencies of $\sim1.67$ mHz and $\sim2.21$ mHz, and correspond observational IDs of 915390701 and 0931791401, respectively. We first fitted the lag-energy spectra using a model comprising a soft-thermal component and a hard-logarithmic component ($A_{\rm ln}+K\ln{[E/E_0]}$). Next, we performed joint fits of the time-averaged spectra, QPO RMS- and lag-energy spectra with the time-dependent Comptonization model \textsc{vkompthdk} \citep{2022MNRAS.515.2099B}. $A_{\rm ln}$ and $A_{\rm bb}$ represent the lag amplitudes of the hard-logarithmic and soft-thermal components, respectively, $K$ is the logarithmic coefficient, and $kT_{\rm bb}$ is the temperature of the thermal lag component. The reference energy for the hard-logarithmic component, $E_0$, is set to $kT_{\rm bb}$. In the dependent Comptonization model, $kT_{\rm s}$ is the seed photon temperature of the oscillating Comptonization component, $L$ is the corona size, and $\eta$ is the reprocessing fraction. Parameter uncertainties are given at the $1\sigma$ confidence level.}
	\label{tab:1} % give each table a logical label name
	\begin{tabular}{lccc} % four columns, alignment for each
		\\
\hline
Model & Parameter & Observation 1 & Observation 2 \\
\hline
Blackbody & $kT_{\rm bb}$ (keV) & $0.176\pm0.032$ & $0.162^{+0.044}_{-0.039}$  \\
 & $A_{\rm bb}$ (s) & $121\pm14$ & $101\pm17$ \\
  & $A_{\rm bb}$ ($R_{\rm g}/c$) & $10.4\pm1.2$ & $8.7\pm1.5$ \\
 \hline
Logarithm & $K$ & $0.25\pm0.13$ & $0.31^{+0.19}_{-0.18}$ \\
 & $A_{\rm ln}$ (s) & $-139\pm33$ & $-121\pm36$\\
\hline
Vkompthdk & $kT_{\rm s}$ (keV) & $0.17\pm0.01$ &  $0.26^{+0.05}_{-0.06}$\\
 & $L$ ($R_{\rm g}$) & $7.1^{+1.1}_{-0.8}$ &  $4.0^{+2.1}_{-1.8}$ \\
 & $\eta$ & $0.88^{+0.09}_{-0.10}$ & $0.53^{+0.19}_{-0.12}$ \\
 \hline
	\end{tabular}
\end{table}

\section{Discussion}
\label{sec:4}
To understand the physical origin of the QPO observed in 1ES 1927+654, it is helpful to draw on the extensive body of work on LFQPOs in BHXBs, which are far more abundant. However, a simple linear scaling of the frequency with mass would suggest a black hole mass for 1ES 1927+654 of $\lesssim 5\times10^4\ M_{\odot}$ \citep{2025Natur.638..370M,2025ApJ...981..125L}, which is inconsistent with the black hole mass estimate of $\sim10^6\ M_{\odot}$ \citep{2022ApJ...933...70L}, making no direct analog between the QPO in 1ES 1927+654 and LFQPOs. Nonetheless, there are notable similarities in the radiative properties of the QPO in 1ES 1927+654 and LFQPOs: (1) the increasing QPO RMS amplitude with energy, and (2) the `U'-shaped lag-energy spectra, suggesting that the physical processes responsible for LFQPOs in BHXBs may still be relevant. Given that our observational and simulation results indicate that the QPO in 1ES 1927+654 originates directly in the Comptonization regions, models based solely on the instability or oscillation of the accretion disk \citep{1990PASJ...42...99K,1999PhR...311..259W} are unlikely to explain the QPO in 1ES 1927+654. Furthermore, models assuming disk instabilities propagating into the corona and inducing oscillations \citep{1999A&A...349.1003T} cannot adequately explain the observed QPO thermal lags. Additionally, the observed thermal lag and the black hole mass of 1ES 1927+654 are consistent with the correlation between the two parameters observed in multiple radio-quiet AGNs (Fig. \ref{fig:S5}), in which the powerful jet is quenched and the radio emission is believed to be from the corona through synchrotron processes \citep{2019NatAs...3..387P,2025ApJ...981..125L}, suggesting the observed mHz QPO in 1ES 1927+654 does not originate within the jet. 

Alternatively, instabilities driven directly in the corona, such as resonant oscillations of the constituent plasma induced by magneto-acoustic wave propagation \citep{2010MNRAS.404..738C,2019MNRAS.490.1350B}, could explain the observed QPO in 1ES 1927+654. 
A simple prescription to describe this mechanism could be magneto-acoustic waves driven by external pressures from the outer regions passing across the corona. The frequency of induced resonant oscillation then scales as $f \approx 2\pi A c_{\rm s} / r_{\rm corona}$ \citep{2010MNRAS.404..738C,2019MNRAS.490.1350B}, where $c_{\rm s} = \sqrt{kT_0 / m_{\rm p}}$ is the sound speed, $k$ is the Boltzmann constant, $T_0$ is the electron temperature, $m_{\rm p}$ is the proton mass, $r_{\rm corona}$ is the radial extent of the corona, and $A$ is a factor of order unity. Assuming a corona plasma temperature of $T_0 = 2 \times 10^8K$, derived from spectral analysis of \textit{Nuclear Spectroscopic Telescope ARray} \citep{2025ApJ...981..125L}, the QPO frequencies of $\sim$1.67 and $\sim$2.21 mHz correspond to $r_{\rm corona}\approx2.5AR_{\rm g}$ and $1.8AR_{\rm g}$, respectively, which are in good agreement with the small corona sizes inferred from the frequency-independent modeling of RMS- and lag-energy spectra (Table \ref{tab:1}). The factor $A$ includes contributions from the turnaround time at each end of the corona and other physical processes which could affect the oscillation frequency \citep{1983SoPh...88..179E,2019MNRAS.490.1350B}. The values of $A$ required for $r_{\rm corona}$ to match the corona sizes inferred from the time-dependent Comptonization model are $\sim2.8$ and $\sim2.2$ for the two observations, respectively. Furthermore, the simultaneous appearance of a newly launched radio jet and the mHz QPO suggests a potential connection between these two phenomena \citep{2025ApJ...979L...2M,2025ApJ...981..125L}. The inward-dragged poloidal magnetic fields that drive jet launching may also serve as the external source triggering the propagating magnetoacoustic waves in the corona, then produce the X-ray QPO.

We notice that from the relation between the black hole mass and QPO frequency, the QPO observed in 1ES 1927+654 shall be a cousin of HFQPOs \citep{2015ApJ...798L...5Z}. However, the rapid evolution in its frequency is unprecedented among HFQPOs observed in both AGNs and BHXBs. The QPO frequency in 1ES 1927+654 shows a significant increase from July 2022 to August 2023, followed by a plateau thereafter (see Fig. \ref{fig:Figure1}B). If this QPO shares a common physical origin with conventional HFQPOs, then HFQPOs in other sources would represent the plateau stage similar to that observed in 1ES 1927+654 post-2024. The frequency-increasing timescale in 1ES 1927+654 is $\sim400$ days. If linked to the viscous timescale in the accretion disk, $t_{\rm vis}$, given by $t_{\rm vis}\sim(\alpha\Omega)^{-1}(H/R)^{-2}\propto\alpha^{-1}M_{\bullet}(R/R_{\rm g})^{3/2}(H/R)^{-2}$, where $\alpha$ is the viscosity parameter, $\Omega$ is the Kepler angular velocity, $R$ is the radius in the disk, and $H$ is the vertical scale height of the disk, the corresponding frequency-increasing timescale in a typical BHXB ($M_\bullet=10M_\odot$) is $\sim250$ s, which is too short for the current  intermittent observational strategies to capture the full frequency evolution. Continued X-ray monitoring is thus crucial to fully compare the QPO properties in 1ES 1927+654 and BHXBs. Under our proposed scenario, the plateau duration should significantly exceed the initial frequency-increasing phase.

It has been suggested that 1ES 1927+654 may be associated with an extreme mass ratio inspiral \citep[EMRI,][]{2025Natur.638..370M,2025ARA&A..63..379K}, which is widely believed to drive quasi-periodic eruptions \citep[QPE,][]{2019Natur.573..381M,2020A&A...636L...2G}, recently detected in the soft X-ray band from the nuclei of a few nearby galaxies. If EMRI models explain the mHz QPO in 1ES 1927+654, a low-mass white dwarf ($\sim0.1M_{\odot}$) could survive within $\lesssim10R_{\rm g}$, filling its Roche lobe at the observed $\sim$10-minute period \citep{2025Natur.638..370M}. This would lead to the emission of gravitational waves at mHz frequencies, detectable by next-generation mHz gravitational-wave detectors, such as the Laser Interferometer Space Antenna (\textit{LISA}), with a signal-to-noise ratio of $\sim$10 \citep{2025Natur.638..370M,2025ARA&A..63..379K}. In EMRI models, the orbiting object passes twice per orbit through the accretion disk of the SMBH on a mildly eccentric, inclined orbit. QPEs are assumed to arise from the emission of an adiabatically expanding, initially optically thick gas cloud expelled from the disk plane during each impact \citep{2023ApJ...957...34L,2023A&A...675A.100F}. However, current EMRI models are essentially related to the thin accretion disk, and remain unclear regarding how the companion interaction with the accretion disk would couple to the corona, which has been convincingly shown to dominate the x-ray QPO radiative properties. If the expelled gas can be somehow channeled into the corona, leading to variability in its optical depth, the observed oscillation in the Comptonization may be explained. However, the soft thermal lags remain difficult to explain in this scenario. In fact, the EMRI model essentially assumes that the oscillation originates from the accretion disk; however, this scenario has been ruled out by our observational and simulation results. Additionally, the EMRI models fail to account for the coronal contraction that coincides with the increase in frequency.

\section{Conclusion}
\label{sec:5}
The SMBH in 1ES 1927+654, exhibiting a high-significance X-ray QPO as an HFQPO cousin, provides a unique laboratory for studying both the origin of the HFQPO and accretion dynamics. The QPO frequency increases rapidly within one year of its initial detection, followed by a stable plateau. For the first time, our novel QPO phase-resolved analysis and modeling reveal an oscillating and contracting compact corona located within a few gravitational radii regions around an SMBH during the frequency-increasing phase. The most likely origin of the observed QPO is plasma resonant oscillations driven by magneto-acoustic waves propagating within the corona. 

\begin{acknowledgments}
%We are grateful to the anonymous referee for constructive comments that helped us improve this paper. 
This research has made use of data obtained from the High Energy Astrophysics Science Archive Research Center (HEASARC), provided by NASA’s Goddard Space Flight Center. This work is supported by the National Key R\&D Program of China (2021YFA0718500) and the National Natural Science Foundation of China (NSFC) under grants, 12025301 and 12333007. This work is partially supported by International Partnership Program of Chinese Academy of Sciences (Grant No.113111KYSB20190020). 
%L. D. Kong is grateful for the financial support provided by the Sino-German (CSC-DAAD) Postdoc Scholarship Program (57251553). P. J. Wang is grateful for the financial support provided by the Sino-German (CSC-DAAD) Postdoc Scholarship Program (57678375).
\end{acknowledgments}





\appendix
\section{Power Spectral Density Analysis}
\label{appendix1}
In this section, we conduct a standard PSD analysis following the procedure from the published work \citep{2025Natur.638..370M}. Using the \textsc{powspec} task within the \textsc{HEASoft} software package 
\footnote{\url{https://heasarc.gsfc.nasa.gov/docs/software/lheasoft/}}, PSDs were computed with a time bin of 20 s in a single segment, yielding a Nyquist frequencies of 0.025 Hz. The PSDs were then normalized to units of fractional RMS squared per Hz \citep{1990A&A...230..103B}. We modeled the unbinned PSDs using a function consisting of two Lorentzian components ($L_1$ and $L_2$) and a constant term ($C_0$). $L_1$ represents the mHz QPO at $\sim1$--$2$ mHz, while $L_2$, with its central frequency fixed at zero, accounts for the red noise dominating at frequencies below $\sim$1 mHz. $C_0$ describes the white noise contribution. To obtain the best-fit parameters of the model, we implemented a Bayesian approach. Considering the fact that the unbinned periodogram follows a $\chi^2$ distribution with two degrees of freedom \citep{2010MNRAS.402..307V}, the log likelihood is given by
\begin{equation}\label{eqS1}
    \log{\mathcal{L}}=-\sum_j{\left(\frac{I_j}{S_j}+\log{S_j}\right)},
\end{equation}
where $I_j$ is the observed power, $S_j$ is the model power and $j$ corresponds to the $j$-th frequency bin.  Due to limited photon statistics, PSD modeling was performed only in two energy bands: 0.3--2 keV and 2--10 keV. The resulting energy dependencies of the fractional RMS are presented in Fig.~\ref{fig:Figure2}A. Table \ref{tab:S1} presents detailed QPO parameters obtained from the PSD modeling.

\section{Variational mode extraction analysis}
\label{appendix2}
We employed a novel technique to perform phase-resolved analysis on two \textit{XMM-Newton} observations (February and August 2023, corresponding observational IDs are 0915390701 and 0931791401) that respectively show QPO detections at frequencies of $\sim$1.67 and $\sim$2.21 mHz with $\gtrsim6\sigma$ significance \citep{2025Natur.638..370M}. Due to the short timescale variations in the recurrence time and amplitude of QPO variability, period folding for phase-resolved analysis is unsuitable. However, the Hilbert-Huang transform (HHT), proposed as an adaptive data analysis method, provides a powerful tool for analyzing signals with nonstationary periodicity \citep{1998RSPSA.454..903H}. The HHT analysis consists of two main steps: mode decomposition, aiming to decompose a signal into several intrinsic mode functions (IMFs), and the Hilbert spectral analysis, enabling the extraction of phase functions for specific IMFs, such as the QPO component \citep{2014ApJ...788...31H,2021MNRAS.500.2475J,2023ApJ...957...84S,2024ApJ...965L...7S,2024ApJ...973...59S}. In the HHT analysis, the empirical mode decomposition (EMD), an algorithm based on interpolation of local extrema, serves as the original mode decomposition technique \citep{1998RSPSA.454..903H}. However, EMD has a significant drawback: the issue of mode mixing, where components of significantly different scales may appear within a single IMF, or a coherent signal can fragment into separate parts appearing in multiple IMFs \citep{2008RvGeo..46.2006H}. As an advanced mode decomposition method, variational mode decomposition (VMD) theoretically mitigates mode mixing by decomposing the signal into a sum of IMFs with analytically determined limited central frequency and bandwidth \citep{2014ITSP...62..531D}. Many studies of LFQPOs observed in BHXBs have demonstrated that VMD is particularly advantageous for decomposing QPO components \citep{2023ApJ...957...84S,2024ApJ...961L..42Z,2024ApJ...965L...7S,2024A&A...692A.117D}. 

The VMD algorithm requires initially setting two parameters: the total number of modes, $K$, and the weighting factor, $\alpha$. One notable challenge with VMD is that the number of modes $K$ must be predetermined before execution. Setting an excessively high number of modes leads to spurious modes appearing, whereas a low number results in mode mixing \citep{2014ITSP...62..531D}. Additionally, in the QPO analysis and numerous other applications, identifying a specific mode within a signal is often the primary objective. In such situations, both the VMD and EMD-based methods impose unnecessarily high computational burdens, as VMD simultaneously extracts all modes, and EMD does so sequentially. To overcome these drawbacks, a technique termed variational mode extraction (VME), built upon principles similar to VMD, was proposed \citep{2018IJBHI...22.1059N}. Utilizing the same conceptual foundation as VMD, VME benefits from a robust mathematical framework. In our VME analysis, we regard the extraction of the QPO as a specific mode extraction problem from the \textit{XMM-Newton} 2--10 keV light curves. Therefore, VME directly seeks a particular mode, and its convergence is unaffected by errors elsewhere, leading to faster convergence and reduced computational demands compared to VMD. These properties, combined with not requiring the number of modes to be set, make VME highly effective in our QPO analysis.

The performance of VME depends primarily on two parameters: the initial value $f_{\rm d}$, representing the approximate central frequency of the target mode, and the parameter $\alpha$, controlling the bandwidth of this mode. For $f_{\rm d}$, we initially set it to the QPO frequency derived from PSD modeling using Lorentzian functions. Regarding $\alpha$, we evaluated reasonable values using two approaches. The simpler approach (Approach 1) involves comparing the full width at half maximum (FWHM) of the QPO Lorentzian to that of the extracted VME mode PSD, seeking a VME mode whose FWHM closely matches that of the QPO Lorentzian. To achieve this, we initially employed various values of $\alpha$ to extract QPO mode from the light curve in the $2$--$10$ keV energy range, determined the corresponding PSD FWHM of the mode, and established the relationship between FWHM and $\alpha$ (see Fig.~\ref{fig:S1}). As shown, the FWHM decreases monotonically with increasing $\alpha$, allowing us to estimate reasonable $\alpha$ values as 9078 and 6895 for the two observations, respectively. The alternative approach (Approach 2) employs the Bayesian PSD modeling, in which we used the log likelihood given by Eq.~\ref{eqS1}, but with the model power replaced by a sum of $L_2$, $C_0$, and the PSD of the extracted QPO component from VME ($P_{\rm VME}$), where $P_{\rm VME}$ includes a parameter $\alpha$. Fig.~\ref{fig:S2} displays posterior probability distributions for each parameter across two observations. Clearly, this Bayesian approach does not tightly constrain $\alpha$, reflecting that the PSD of the extracted QPO component exhibits limited sensitivity to $\alpha$ once above a certain lower limit. Fig.~\ref{fig:S3} presents the light curves and PSDs of the extracted QPO component sampled from posterior distributions obtained via nested sampling. Evidently, the VME results remain stable for values of $\alpha$ above the Bayesian-derived lower limit. Additionally, Fig.~\ref{fig:S3} illustrates results using $\alpha$ values obtained from Approach 1, demonstrating good agreement with those derived from Approach 2. Because the Bayesian method does not strongly constrain $\alpha$, we adopted the $\alpha$ values obtained from Approach 1 for all subsequent analyses.

\setcounter{figure}{0}
\renewcommand{\thefigure}{B\arabic{figure}}
\renewcommand{\theHfigure}{B\arabic{figure}}

\begin{figure}
\centering
    \includegraphics[width=0.5\textwidth]{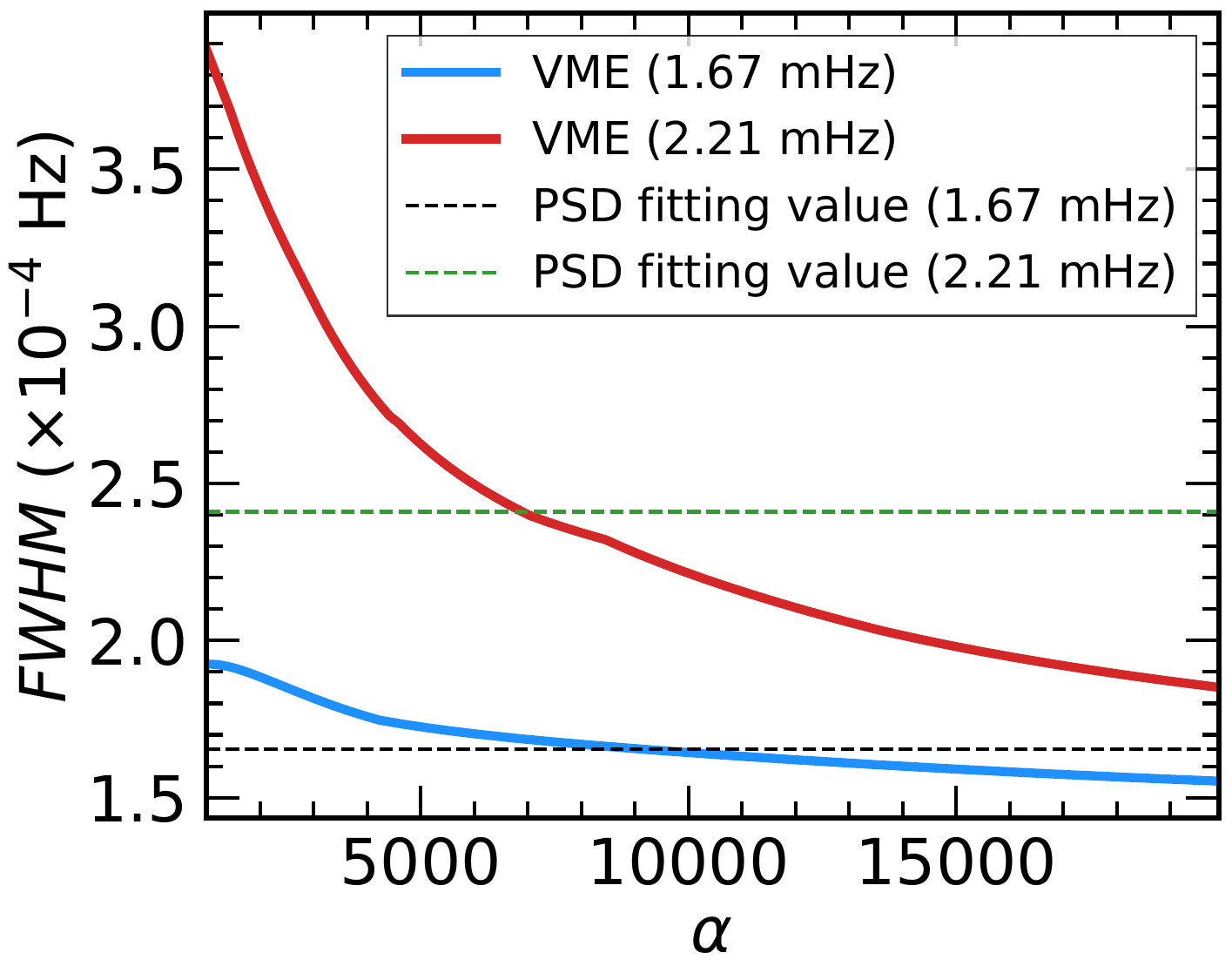}
    \caption{Relationship between PSD FWHM of the extracted QPO mode and the VME $\alpha$ parameter. The horizontal dashed lines represent the FWHM values obtained from the PSD modeling of the mHz QPO with a Lorentzian function. In the VME analysis, the FWHM decreases monotonically with increasing $\alpha$, allowing us to estimate reasonable $\alpha$ values for two observations as 9078 and 6895, respectively. \label{fig:S1}}
\end{figure}

\begin{figure} % Do not use \begin{figure*}
\centering
\begin{minipage}[c]{0.4\linewidth}
\centering
    \includegraphics[width=\linewidth]{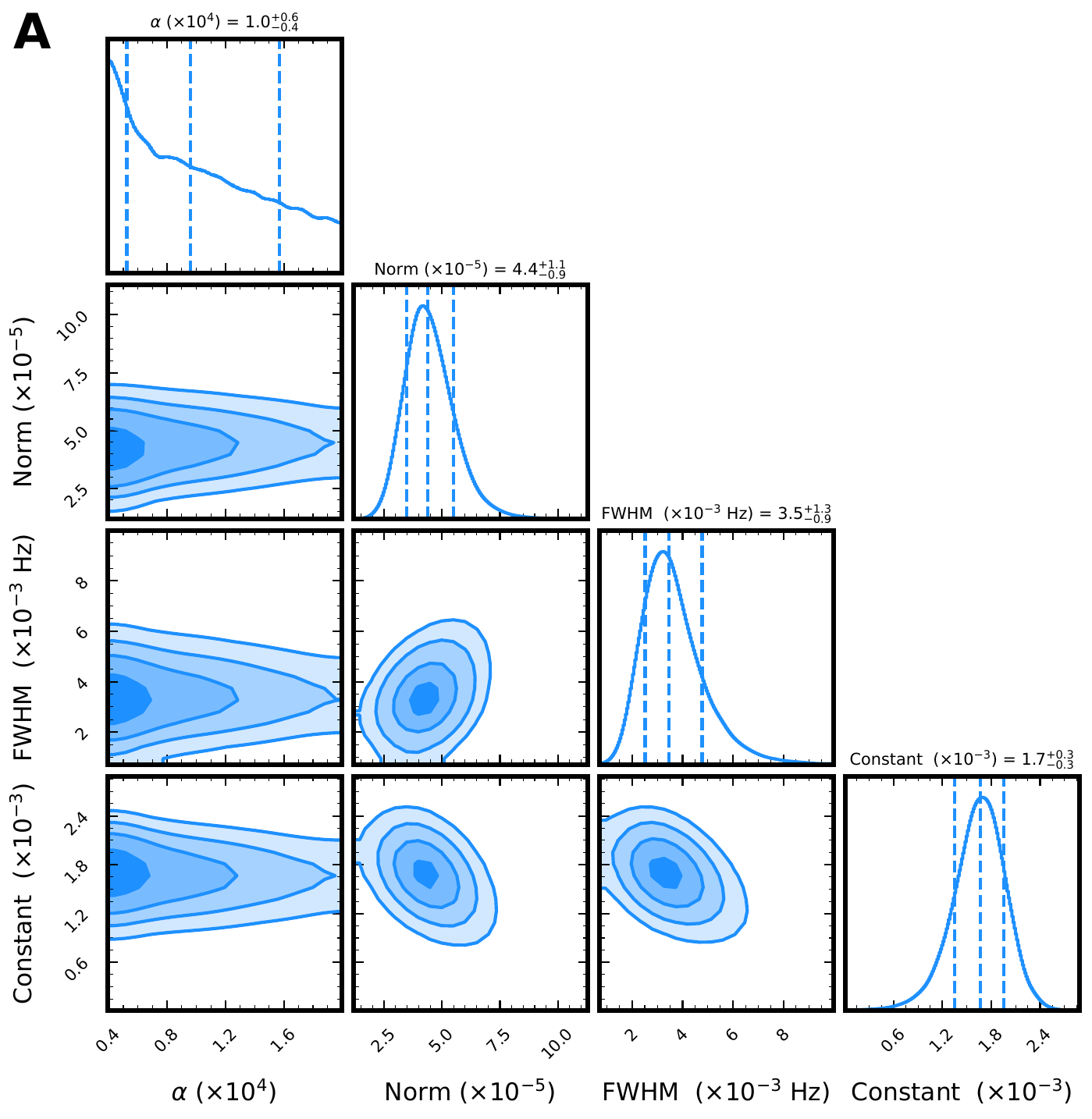}
\end{minipage}
\begin{minipage}[c]{0.4\linewidth}
\centering
    \includegraphics[width=\linewidth]{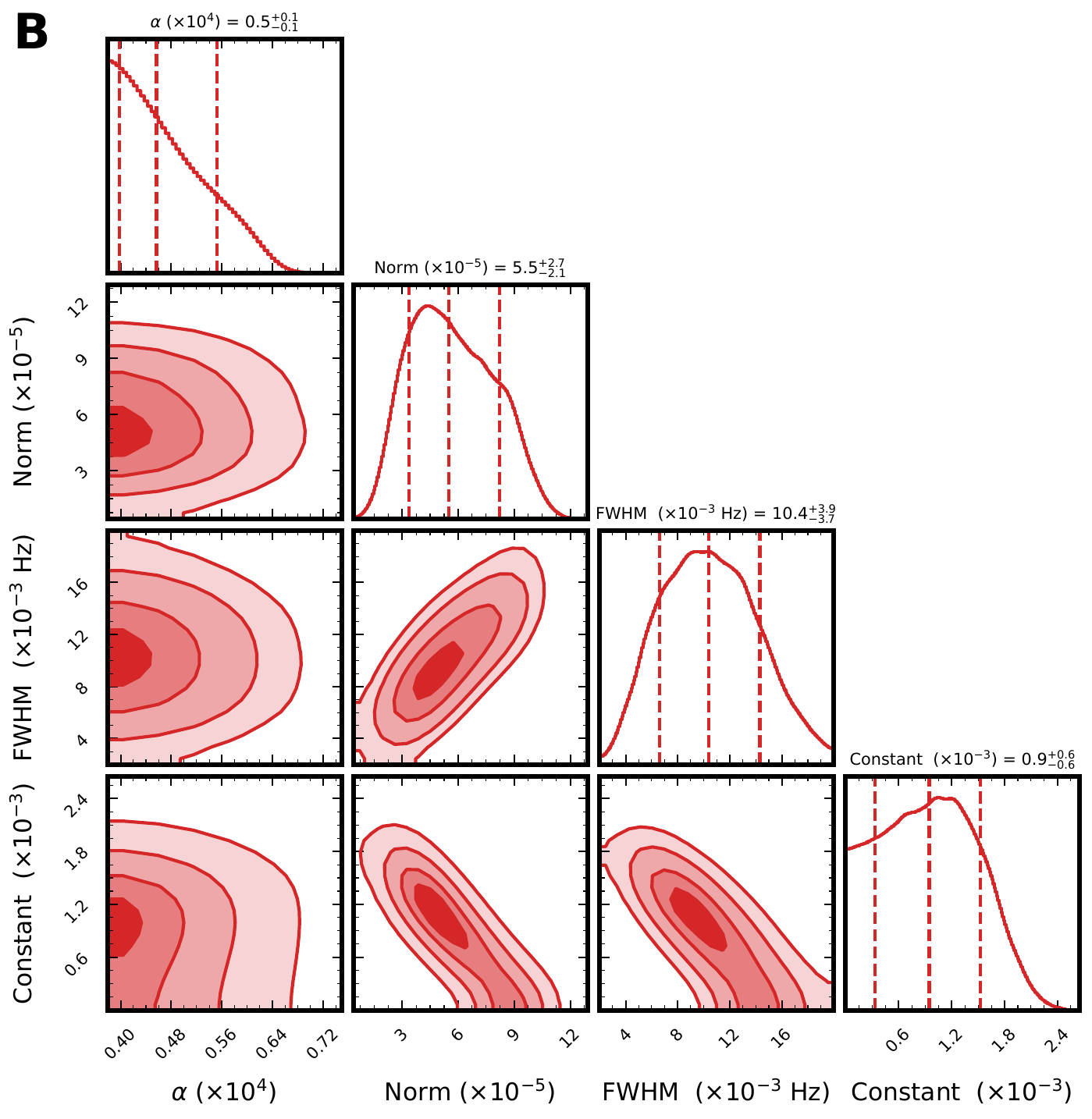}
\end{minipage} 
\caption{Posterior probability distributions for parameters derived from Bayesian modeling to estimate the VME parameter $\alpha$ for two observations. (\textbf{A}) Distributions corresponding to the observation with the 1.67 mHz QPO. (\textbf{B}) Distributions corresponding to the observation with the 2.21 mHz QPO. \label{fig:S2}}  % give each figure a logical label name
\end{figure}

\begin{figure} % Do not use \begin{figure*}
\centering
    \includegraphics[width=0.7\linewidth]{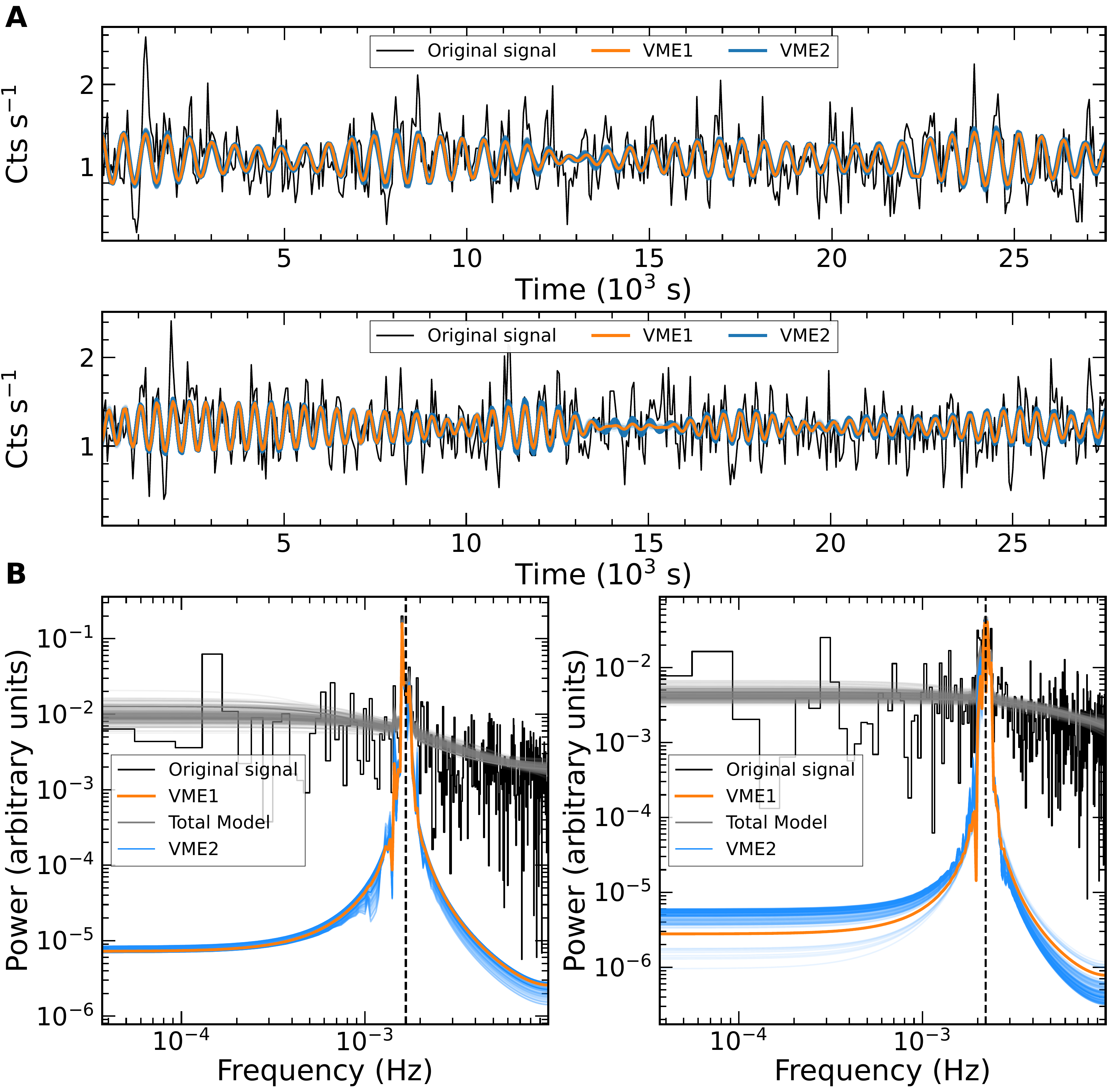}
\caption{QPO light curves and corresponding PSDs derived from VME analysis. (\textbf{A}) Extracted QPO light curves using the VME method. Orange lines show results obtained with the optimal $\alpha$ values determined from Fig.~\ref{fig:S1} (Approach 1). Blue lines represent results from the posterior distributions derived through Bayesian approach (Approach 2). Black lines indicate the original \textit{XMM-Newton} 2--10 keV light curves. (\textbf{B}) PSDs corresponding to the extracted QPO light curves. The colors of the curves match those in panel (\textbf{A}). Original PSDs are shown in black, while the gray curves represent the total PSD model (consisting of $L_2$, $C_0$, and $P_{\rm VME}$ components) plotted from the posterior distributions of the Bayesian approach. \label{fig:S3}}
\end{figure}

\section{Phase-resolved analysis}
\label{appendix3}
For the extracted intrinsic QPO light curves, $Q(t)$, physically meaningful phase information can be obtained using the Hilbert transform. The Hilbert transform of $Q(t)$ is defined as
\begin{equation}
\mathcal{H}\left[Q(t)\right]=\frac{1}{\pi}{\rm pv}\int\frac{Q(\tau)}{t-\tau}d\tau,
\end{equation}
where pv denotes the Cauchy principal value. Using this transform, the analytic signal corresponding to $Q(t)$ can be constructed as
\begin{equation}
    Q_{\rm A}(t)=Q(t)+j\mathcal{H}\left[Q(t)\right],
\end{equation}
where $j$ is an imaginary unit. The instantaneous phase function, $\phi(t)$, can then be calculated by
\begin{equation}
    \phi(t)=\arctan{\left\{\frac{\mathcal{H}\left[Q(t)\right]}{Q(t)}\right\}}.
\end{equation}
Because the mHz QPO signal is strongest in the 2--10 keV energy band, both the VME extraction and the subsequent Hilbert phase computation were performed within this energy range. Fig.~\ref{fig:S4} plots count rate as a function of the 2--10 keV QPO phase (phase-folded light curve) for three representative energy bands. To investigate the energy dependence of QPO amplitude and lag, we performed sinusoidal fits to the phase-folded light curves in eight distinct energy bands (representative best-fit models are shown in Fig. \ref{fig:S4}). The sinusoidal model used for fits is
\begin{equation}
    \mathcal{F}(\phi, E) = R(E)\times\left\{1+\sqrt{2}\sigma_{\rm RMS}(E)\cos{\left[2\pi\left(\phi-\psi(E)-\frac{1}{2}\right)\right]}\right\},
\end{equation}
where $\mathcal{F}(\phi, E)$ represents the phase-folded light curve at energy $E$, QPO phase $\phi$ is expressed in units of QPO cycles, $R(E)$ is the average count rate, $\sigma_{\rm RMS}(E)$ is the QPO RMS amplitude, and $\psi(E)$ represents the QPO phase lag at energy $E$ relative to the 2--10 keV reference band. After determining $\psi(E)$, any energy band can be chosen as the reference. In Fig.~\ref{fig:Figure2}B, we adopted the 0.3--0.5 keV band as the reference energy band. 

We observe significant soft lags between energy bands below $\sim 1$ keV and higher energies. For consistency with the previous time-lag analysis in AGNs \citep{2013MNRAS.431.2441D}, we additionally calculated QPO time lags between the 0.3--1 and 1--5 keV energy bands, and compared them with soft reverberation lags found in other non-QPO AGNs. We find that the observed QPO soft lag values and the black hole mass of 1ES 1927+654 are consistent with the correlation between the two parameters observed in multiple radio-quiet AGNs \citep{2013MNRAS.431.2441D} (see Fig.~\ref{fig:S5}).

Using the well-defined phase functions obtained from the analysis described above, we extracted \textit{XMM-Newton} EPIC-PN spectra from six distinct phase bins for each of the two QPO observations, enabling phase-resolved spectral analysis. We modeled these spectra using a phenomenological model: \textsc{tbabs}$\times$\textsc{ztbabs}$\times$(\textsc{zpowerlw}+\textsc{zbbody}). The \textsc{tbabs} component represents the Milky Way interstellar absorption, with the column density ($N_{\rm H}$) fixed at $6.42\times10^{20}\ {\rm cm}^{-2}$ \citep{2005A&A...440..775K}. The \textsc{ztbabs} component describes intrinsic absorption from the host galaxy, with the column density treated as a free parameter during fitting. The \textsc{zpowerlw} and \textsc{zbbody} components account for Comptonization and soft excess emission, respectively. To avoid degeneracy among spectral parameters, the intrinsic absorption column density from the host galaxy was tied across all phase-resolved spectra. Joint fitting of the six phase-resolved spectra using this model yielded acceptable $\chi^2/{\rm d.o.f.}$ values of 477.77/454 and 484.63/459 for the two observations, respectively. We further used the \textsc{cflux} model to calculate unabsorbed fluxes in the 0.3--2 keV ($F_{\rm 0.3-2\ keV}$) and 2--10 keV ($F_{\rm 2-10\ keV}$) energy bands and derived the ratio of these fluxes to determine the hardness ratio ($HR=F_{\rm 2-10\ keV}/F_{\rm 0.3-2\ keV}$). Fig.~\ref{fig:S6} presents the variability of spectral parameters as a function of the mHz QPO phase, based on jointly fitting results from phase-resolved spectra. The source spectrum clearly becomes harder during the QPO peaks, as evidenced by the anti-correlation between the spectral index and flux, and the synchronous modulation of hardness ratio with flux.

\setcounter{figure}{0}
\renewcommand{\thefigure}{C\arabic{figure}}
\renewcommand{\theHfigure}{C\arabic{figure}}

\begin{figure} % Do not use \begin{figure*}
\centering
    \includegraphics[width=0.75\linewidth]{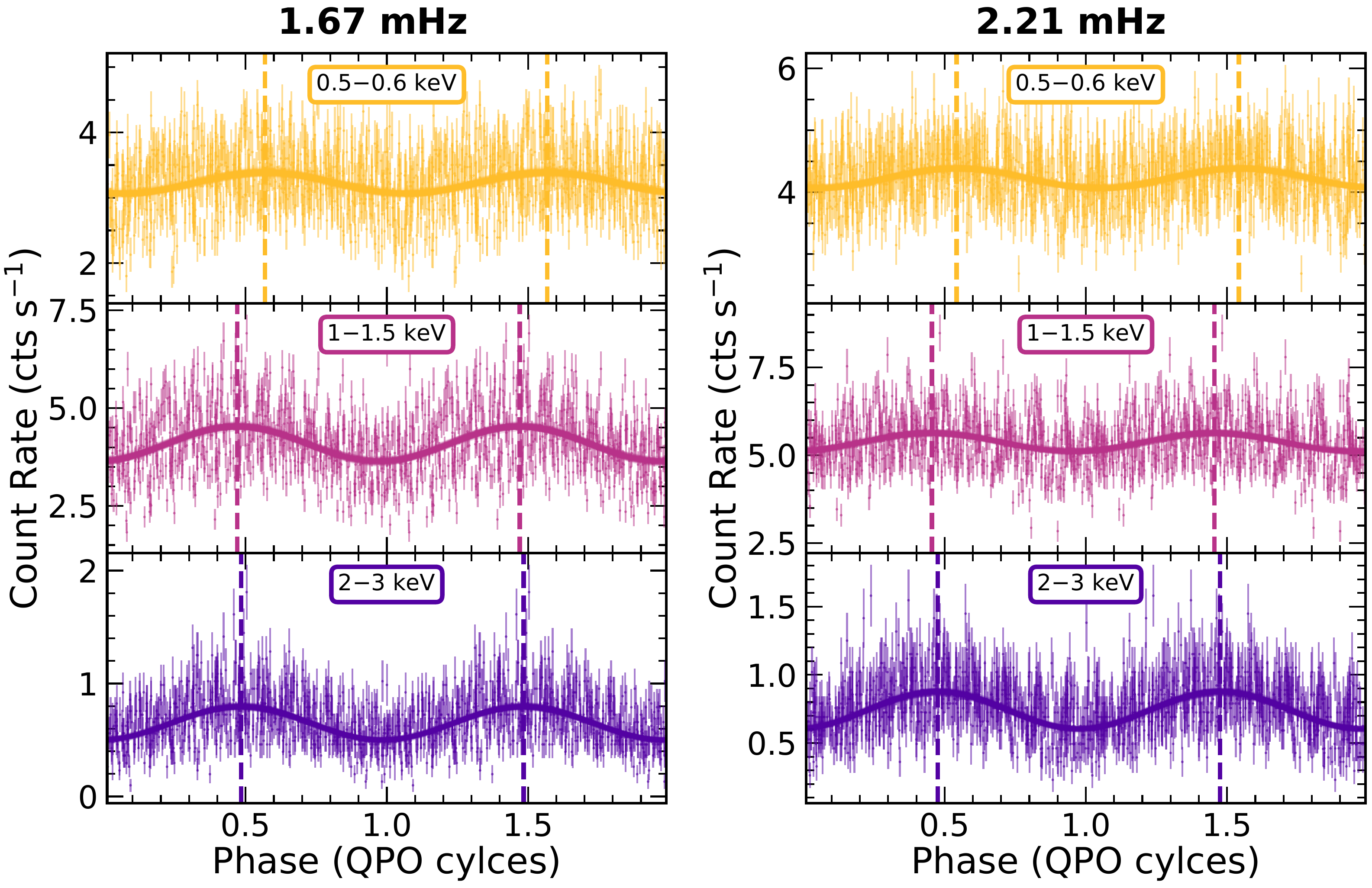}
\caption{Phase-folded QPO light curves in three representative energy bands, fitted with a sinusoidal model. Data points with error bars represent observed count rates in each time bin; solid lines indicate the best-fit sinusoidal models. Vertical dashed lines mark the best-fit peak phases of the QPO waveform. \label{fig:S4}}
\end{figure}

\begin{figure} % Do not use \begin{figure*}
\centering
    \includegraphics[width=0.75\linewidth]{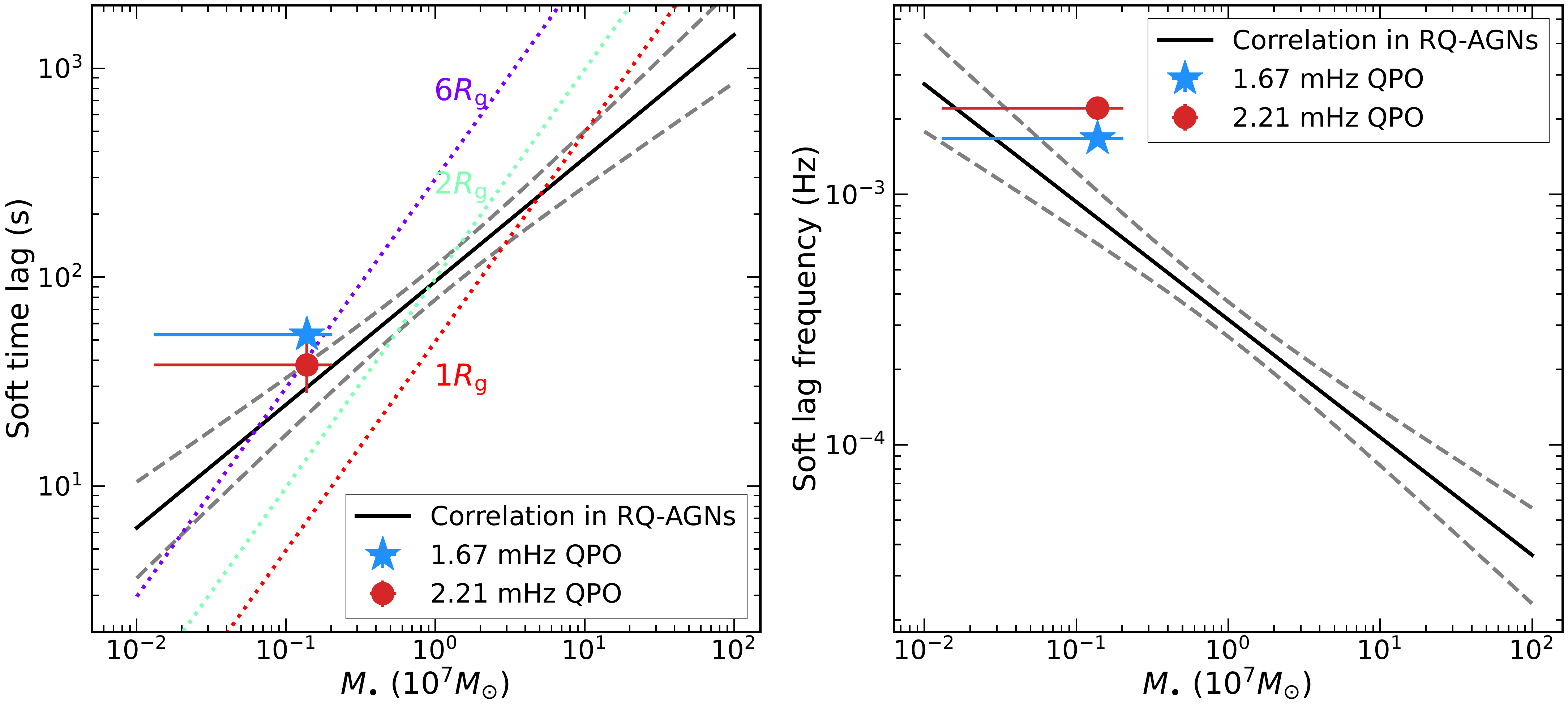}
\caption{Soft lag amplitude versus black hole mass and soft lag frequency versus black hole mass trends. Data points with error bars represent the observational results of the mHz QPO in 1ES 1927+654. The best-fit linear models (in log-log space) for the correlations observed in multiple radio-quiet AGNs (RQ-AGNs) \citep{2013MNRAS.431.2441D}, along with the combined 1$\sigma$ error on the slope and normalization, are overlaid as continuous and dashed lines. The plots of the models and the corresponding 1$\sigma$ confidence intervals are directly taken from the public work \citep{2013MNRAS.431.2441D}. The dotted lines in the left panel represent the light-crossing time at $1R_{\rm g}$, $2R_{\rm g}$, and $6R_{\rm g}$ as a function of black hole mass. The plotted soft lag frequency of 1ES 1927+654 is set to the observed QPO frequency. \label{fig:S5}}
\end{figure}

\begin{figure} % Do not use \begin{figure*}
\centering
    \includegraphics[width=0.75\linewidth]{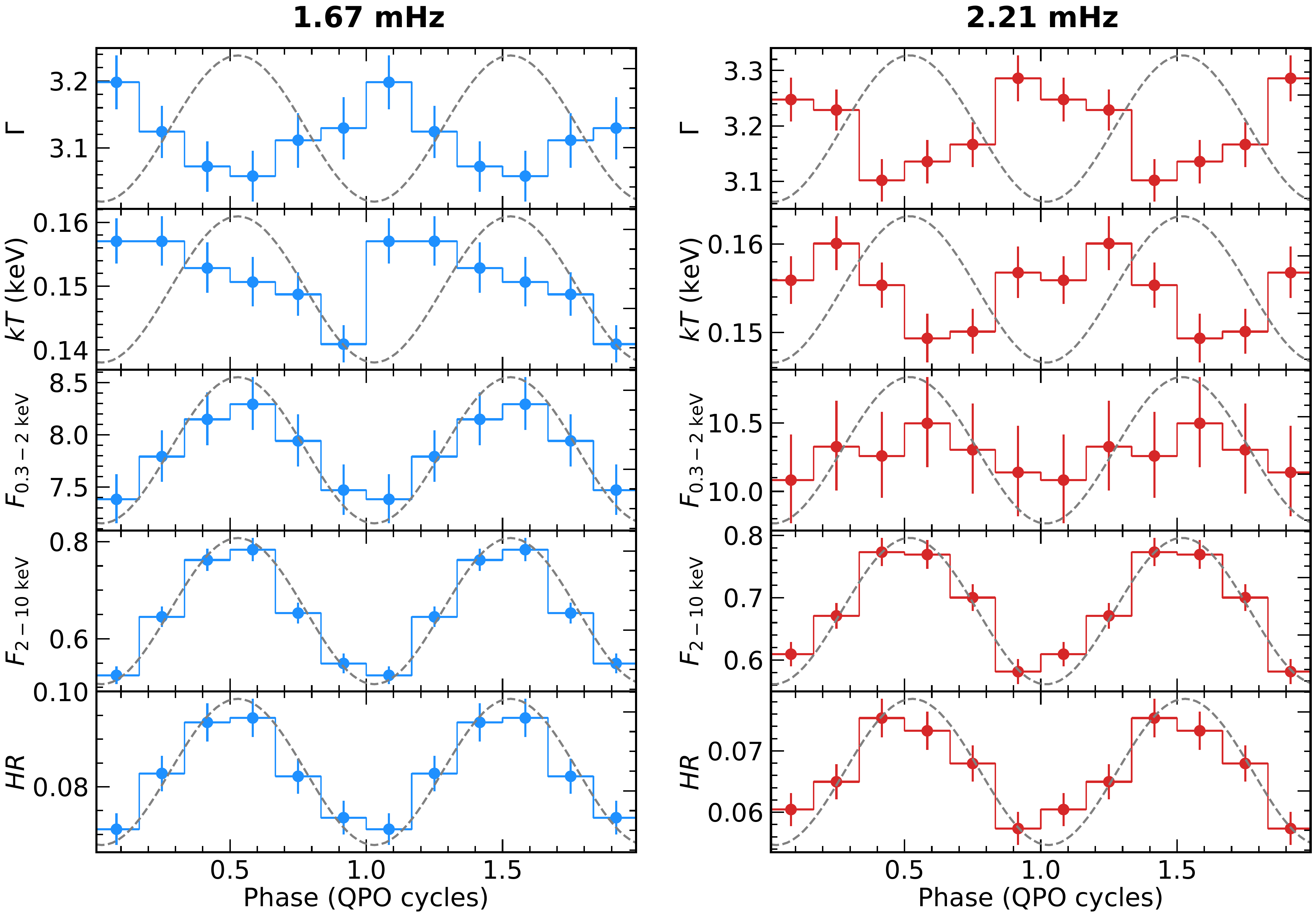}
\caption{Variability of spectral parameters over the mHz QPO cycle derived from phase-resolved spectral analysis.
From top to bottom panels: spectral index ($\Gamma$), blackbody temperature ($kT$), unabsorbed model fluxes (in units of $10^{-11}\rm erg\ s^{-1}\ cm^{-2}$) for the energy ranges 0.3--2 keV and 2--10 keV, and hardness ratio (flux ratio between these two bands). The phase-folded QPO light curve in the 2--10 keV energy band is plotted in each panel as a gray dashed line for reference. \label{fig:S6}}
\end{figure}

\section{Joint fitting of the time-averaged spectra, RMS- and lag-energy spectra}
\label{appendix4}
In this section, we jointly fitted the time-averaged energy spectra of 1ES 1927+654, along with the RMS- and lag-energy spectra of the mHz QPO obtained from the phase-resolved analysis, using the single-component Comptonization model \textsc{vkompthdk} \citep{2022MNRAS.515.2099B}. The \textsc{vkompthdk} model assumes that the seed photons originate from a geometrically thin, optically thick accretion disk \citep{1973A&A....24..337S}, with a temperature $kT_{\rm s}$ at the innermost radius. These photons undergo inverse-Compton scattering within a corona, modeled as a spherically symmetric homogeneous plasma with size $L$ and electron temperature $kT_{\rm e}$. The corona is filled with highly energetic electrons, uniformly distributed with a number density $n_{\rm e}$, and is kept in thermal equilibrium by external heating. The model also incorporates a feedback process, where a fraction of the photons upscattered in the corona are redirected back onto the accretion disk. The feedback fraction, $0 \leq \eta \leq 1$, represents the fraction of the disk flux resulting from coronal irradiation. This model does not assume a specific dynamic origin for the QPO, but instead treats it as small oscillations of the spectrum around the time-averaged one. These spectral fluctuations are attributed to perturbations in the electron temperature $kT_{\rm e}$, via feedback, of the seed photon source temperature $kT_{\rm s}$, as a result of an oscillating external heating rate $\delta \dot{H}_{\rm ext}$ \citep{2022MNRAS.515.2099B}. The key parameters of the model include the seed photon source temperature $kT_{\rm s}$, the electron temperature $kT_{\rm e}$, the power-law photon index $\Gamma$, the size of the corona $L$, the feedback fraction $\eta$, and the variation of the external heating rate $\delta \dot{H}_{\rm ext}$. 

In the joint fit, we modeled the time-averaged spectra using \textsc{tbabs} $\times$ \textsc{ztbabs} $\times$ (\textsc{nthcomp} + \textsc{zbbody}), where the \textsc{nthcomp} model represents a thermally Comptonized continuum. Since the fit was performed within the \textit{XMM-Newton} EPIC-PN energy range of 0.3--10 keV, we fixed the electron temperature, $kT_{\rm e}$, at 8.4 keV, based on the recent \textit{Nuclear Spectroscopic Telescope Array} (\textit{NuSTAR}) spectral analysis \citep{2025ApJ...981..125L}. The \textsc{vkompthdk} model was added as an external component to fit the RMS- and lag-energy spectra of the mHz QPO. It is important to note that the steady-state spectrum of \textsc{vkompthdk} matches that of \textsc{nthcomp}, so we initially linked the seed photon source temperature of \textsc{vkompthdk}, $kT_{\rm s}$, to that of the \textsc{nthcomp} model, $kT_{\rm b}$. The initial fits (Case 1) for both observations resulted in $\chi^2/{\rm d.o.f.}$ values of 181.7/133 and 200.73/131, respectively. The $\chi^2$ values and degrees of freedom were primarily influenced by the time-averaged spectrum, with the main residuals arising from the fit to the RMS- and lag-energy spectra. Allowing $kT_{\rm s}$ and $kT_{\rm b}$ to vary independently (Case 2) significantly improved the fit to the RMS and phase-lag spectra. Table \ref{tab:S2} summarizes the joint fitting results for both cases. Fig.~\ref{fig:S7} presents the best-fit results of \textsc{vkompthdk} model of Case 2 for the 1.67-mHz QPO observation. The plot clearly shows that the \textsc{vkompthdk} model provides a good description of the RMS- and lag-energy spectra of the mHz QPO. The derived corona sizes are $1.45^{+0.22}_{-0.16}\times10^7$ and $0.82^{+0.42}_{-0.37}\times10^7$ km, which correspond to $7.1^{+1.1}_{-0.8}$ and $4.0^{+2.1}_{-1.8}R{\rm g}$ for the two observations, respectively, with a black hole mass of $1.38\times10^{6}$ $M_{\odot}$, and a relatively large feedback fraction, $\eta \gtrsim 0.5$. Moreover, the best-fit corona size shows a decrease as the QPO frequency increases. The averaged seed photon source temperature derived from the \textsc{vkompthdk} model is $kT_{\rm s} \approx 0.2$ keV, significantly higher than that of the time-averaged Comptonization continuum, $kT_{\rm b} \approx 50$ eV. Notably, the best-fit values of $kT_{\rm b}$ are close to the accretion disk temperature obtained from broad-band spectral energy distribution (SED) fitting \citep{2024ApJ...975...50L}, while those of $kT_{\rm s}$ are consistent with the temperature of the soft excess blackbody component ($kT \approx 0.16$ keV).

\setcounter{figure}{0}
\renewcommand{\thefigure}{D\arabic{figure}}
\renewcommand{\theHfigure}{D\arabic{figure}}

\setcounter{table}{0}
\renewcommand{\thetable}{D\arabic{table}}
\renewcommand{\theHtable}{D\arabic{table}}

\begin{table} % Do not use \begin{table*}
	\centering
	% Captions go above tables
	\caption{Best-fit parameters of the joint fit to the time-averaged energy spectrum of 1ES 1927+654, and the RMS- and lag-energy spectra of the mHz QPO, using the single-component Comptonization model \textsc{vkompthdk}.  Uncertainties are given at the 68 per cent ($1\sigma$) confidence level.}
	\label{tab:S2} % give each table a logical label name
	\begin{tabular}{lccccc} % four columns, alignment for each
		\\
\hline
Component & Parameter & 1.67 mHz & 1.67 mHz & 2.21 mHz & 2.21 mHz\\
 &  & (Case 1) & (Case 2) & (Case 1) & (Case 2)\\
\hline
Ztbabs & $N_{\rm H}$ ($10^{20}$ cm$^{-2}$) & $0.81^{+0.24}_{-0.30}$ & $3.14^{+0.62}_{-0.36}$ & $3.07^{+0.64}_{-0.50}$ &  $3.97^{+0.58}_{-0.55}$ \\
Zbbody & $kT$ (keV) & $0.160^{+0.001}_{-0.002}$ & $0.154^{+0.002}_{-0.003}$ & $0.159^{+0.001}_{-0.002}$ & $0.156\pm0.002$ \\
 & $N_{\rm bb}$ ($10^{-4}$) & $2.21^{+0.09}_{-0.07}$ & $3.14^{+0.15}_{-0.10}$ & $4.43^{+0.29}_{-0.19}$ & $4.88^{+0.23}_{-0.19}$ \\
Nthcomp & $\Gamma$ & $3.02\pm0.02$ & $3.00^{+0.03}_{-0.02}$ & $3.10\pm0.03$ & $3.10^{+0.03}_{-0.02}$ \\
 & $kT_{\rm b}$ (eV) & $=kT_{\rm s}$ & $55^{+7}_{-43}$ & $=kT_{\rm s}$ & $62^{+9}_{-34}$ \\
 & $N_{\rm comp}$ ($10^{-2}$) & $1.12\pm0.02$ & $1.16^{+0.03}_{-0.02}$ & $1.31\pm0.03$ & $1.32\pm0.03$ \\
 Vkompthdk & $kT_{\rm s}$ (keV) & $0.11^{+0.05}_{-0.02}$ & $0.17\pm0.01$ & $0.08\pm0.01$ & $0.26^{+0.05}_{-0.06}$\\
 & $L$ ($R_{\rm g}$) & $6.0^{+0.2}_{-0.4}$ & $7.1^{+1.1}_{-0.8}$ & $3.7^{+0.6}_{-1.3}$ & $4.0^{+2.1}_{-1.8}$ \\
 & $\eta$ & $1.0^{c}_{-0.1}$ & $0.88^{+0.09}_{-0.10}$ & $1.0^c_{-0.5}$ & $0.53^{+0.19}_{-0.12}$ \\
 & $\delta\dot{H}_{\rm ext}$ & $393^{+32}_{-78}$ & $156^{+85}_{-46}$ & $231^{+127}_{-112}$ & $8^{+42}_{-2}$ \\
\hline
  & $\chi^2$/d.o.f. & 181.7/133 & 141.1/132 & 200.7/131 & 164.6/130 \\
 \hline
	\end{tabular}
\end{table}

\begin{figure} % Do not use \begin{figure*}
\centering
    \includegraphics[width=0.95\linewidth]{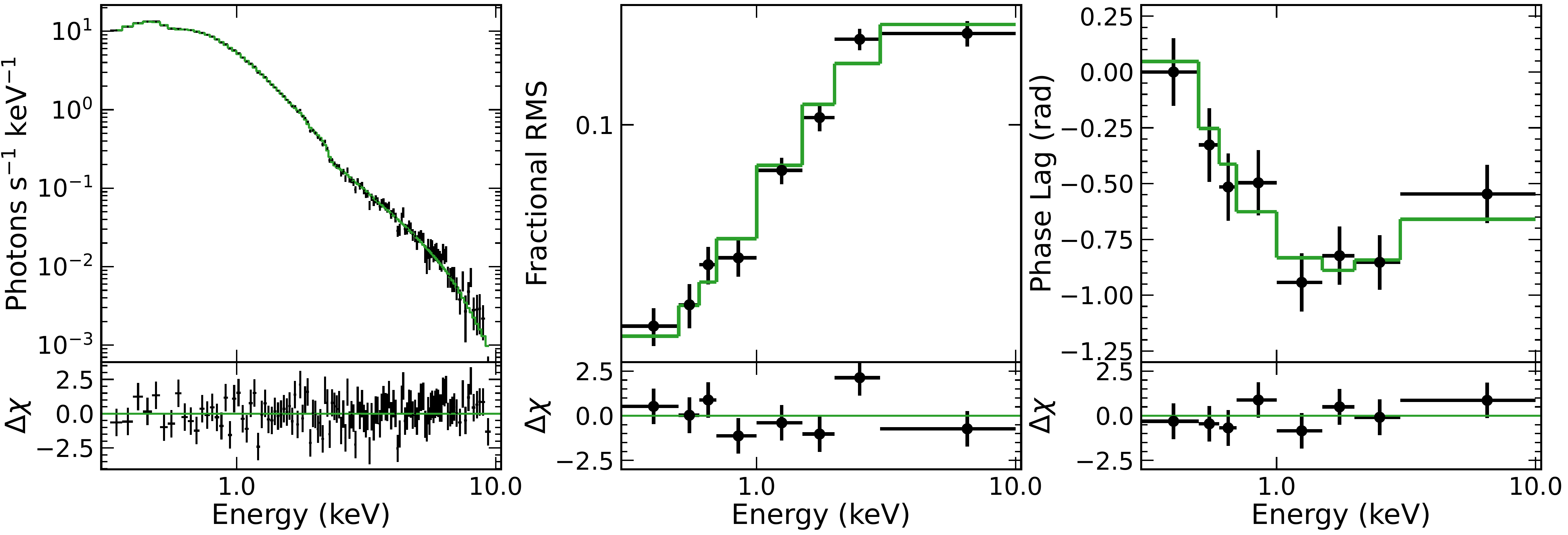}
\caption{Joint fitting of the time-averaged energy spectrum, the QPO RMS- and lag-energy spectra for the 1.67-mHz-QPO observation (obsID 0915390701). In the middle and right panels, the solid green lines represent the best-fitting model with \textsc{vkompthdk}. Bottom panels: residuals.  \label{fig:S7}}
\end{figure}

\section{Monte Carlo simulation of an oscillating corona}
\label{appendix5}
In this section, we performed a numerical simulation using the Monte Carlo radiative transfer code \textsc{monk} \citep{2019ApJ...875..148Z} to demonstrate that the observed QPO spectral properties can be explained by oscillations in the coronal temperature and/or optical depth, rather than by oscillations in the accretion rate of the accretion disk. In \textsc{monk}, optical-UV seed photons are generated according to the disk emissivity. These photons are ray-traced along null geodesics in Kerr spacetime, and Compton scattered after reaching the corona regions assuming the Klein-Nishina cross-section. The photon propagation terminates when the photons either enter the event horizon, hit the disk, or reach infinity. By counting the latter photons, the energy and polarization spectra can be constructed. Since polarimetric analysis was not included in our work, we disabled the polarimetric calculation in \textsc{monk}. 

The primary input parameters for \textsc{monk} are the black hole mass and spin, the accretion rate, and the physical (optical depth and temperature) and geometrical parameters of the corona. For our simulations, we set the black hole mass to $1.38\times10^6M_{\odot}$. For simplicity, we modeled the accretion flow geometry in \textsc{monk} by assuming that the inner regions of the standard thin disk are covered by a slab corona. The inner edges of both the corona and disk are set to extend down to the innermost stable circular orbit (ISCO), with the radial extent and vertical thickness of the corona set to $5 R_{\rm g}$, consistent with the best-fit values from the jointly fitting results (see Table~\ref{tab:S2}). By introducing a sinusoidal modulation in Eddington ratio of the accretion rate $\dot{m}\equiv\dot{M}/\dot{M}_{\rm Edd}$ (where $\dot{M}$ is the mass accretion rate and $\dot{M}_{\rm Edd}$ is the Eddington limit of the mass accretion rate), the optical depth $\tau$, and the temperature $kT_{\rm e}$, we simulated three types of QPO spectral variability corresponding to modulations in these three parameters. The time-averaged $\dot{m}_0$ was inferred to be $\sim$0.57 using equation (11) from the public work \citep{2024ApJ...975...50L}, the black hole spin was set to 0.998, and the time-averaged coronal temperature $kT_{\rm e0}$ was set to 8.4 keV based on recent \textit{NuSTAR} spectral analysis \citep{2025ApJ...981..125L}. The time-averaged coronal optical depth $\tau_0$ was set to 1.5 to match the spectral index of the Comptonization spectra from \textsc{monk} with our \textit{XMM-Newton} spectral analysis ($\Gamma \sim 3.1$). The modulation patterns for the three input parameters are: $\dot{m}=\dot{m}_0(1 + Ae^{j\phi_{\rm QPO}})$, $\tau=\tau_0(1 + Ae^{j\phi_{\rm QPO}})$, and $kT_{\rm e}=kT_{\rm e0}(1 + Ae^{j\phi_{\rm QPO}})$, where $A = 0.05$ is the assumed amplitude for each parameter, $j$ is the imaginary unit, and $\phi_{\rm QPO}$ is the QPO phase, ranging from 0 to $2\pi$. We performed simulations for 10 phase bins from $0$ to $\pi$, since the input parameter values for the phase range $\pi$ to $2\pi$ are identical to those for the first half of the cycle.

For each phase bin, we computed the Comptonization spectrum, the corresponding integral fluxes in the 0.3--10, 0.3--2, and 2--10 keV energy bands, as well as the hardness ratio, defined as the flux ratio between the 2--10 and 0.3--2 keV energy bands. Additionally, using the simulated phase-resolved spectra, we calculated QPO spectra $F_{\rm QPO}(E)$ as:
\begin{equation}
    F_{\rm QPO}(E)=\left\{\sum_{i=1}^N\left[F_i(E)-\bar{F}(E)\right]^2/N\right\}^{1/2},
\end{equation}
where $\bar{F}(E)$ is the time-averaged spectral intensity, $F_i(E)$ is the spectral intensity at the $i$-th phase bin, and $N = 20$ is the total number of phase bins in the QPO cycle. As shown in Fig.~\ref{fig:S8}, when only the accretion rate is modulated, the QPO spectrum closely resembles the time-averaged and phase-resolved spectra, with no significant variability in the hardness ratio. In addition, the modulation in $\dot{m}$ alone does not lead to any noticeable changes in the spectral shape of the Comptonization emission throughout the QPO cycle. These outcomes are inconsistent with our observational results shown in Fig. \ref{fig:Figure3}. However, when the coronal optical depth or temperature are modulated, the QPO spectrum becomes much harder than the time-averaged spectrum above $\sim$1 keV. Both patterns of modulation in optical depth and temperature demonstrate synchronous modulations in the hardness ratio with flux throughout the QPO cycle. Therefore, it is clear that the coronal properties must change during the QPO cycle to produce the observed QPO spectral properties in the X-ray band.

\setcounter{figure}{0}
\renewcommand{\thefigure}{E\arabic{figure}}
\renewcommand{\theHfigure}{E\arabic{figure}}

\begin{figure} % Do not use \begin{figure*}
\centering
    \includegraphics[width=0.85\linewidth]{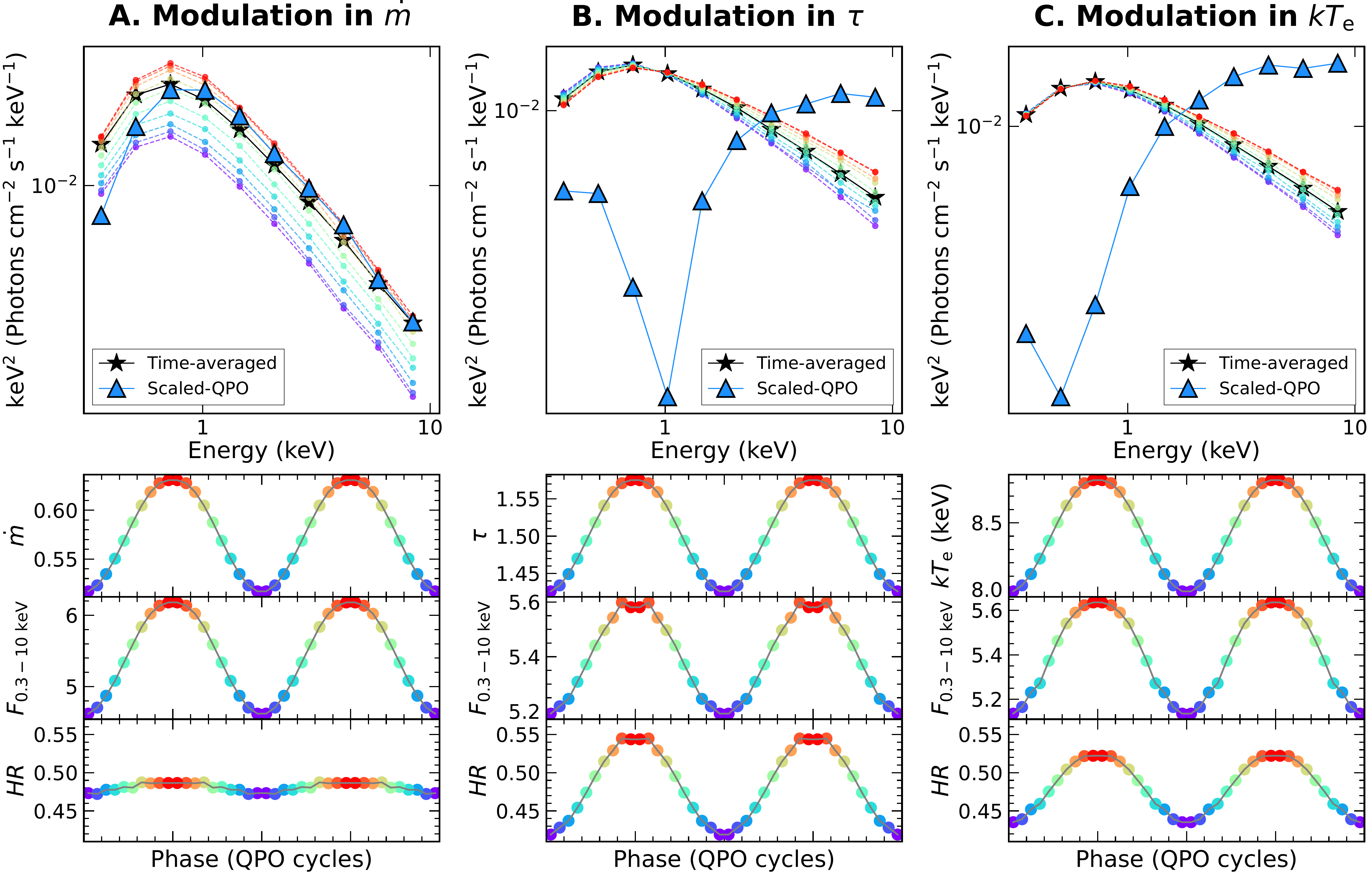}
\caption{Spectral properties of the mHz QPO from a Monte Carlo simulation analysis using the \textsc{monk} code. The top panels present the time-averaged (black stars), representative phase-resolved (colored dots), and QPO spectra (blue triangles, scaled by relevant factors for better comparison) with modulations in the accretion rate $\dot{m}$, coronal optical depth $\tau$, and corona temperature $kT_{\rm e}$, respectively, while the lower panels present the corresponding modulations in parameters during the QPO cycle, where $F_{\rm 0.3-10\ keV}$ represents flux in the energy range of 0.3--10 keV in units of $10^{-11}\rm erg\ s^{-1}\ cm^{-2}$, and $HR$ represents the hardness ratio (defined as flux ratio between the 2--10 keV and 0.3--2 keV energy bands). (\textbf{A}) spectral properties with modulation in $\dot{m}$. (\textbf{B}) spectral properties with modulation in $\tau$. (\textbf{C}) spectral properties with modulation in $kT_{\rm e}$. \label{fig:S8}}
\end{figure}


\bibliography{sample631}{}

\begin{thebibliography}{}
\expandafter\ifx\csname natexlab\endcsname\relax\def\natexlab#1{#1}\fi
\providecommand{\url}[1]{\href{#1}{#1}}
\providecommand{\dodoi}[1]{doi:~\href{http://doi.org/#1}{\nolinkurl{#1}}}
\providecommand{\doeprint}[1]{\href{http://ascl.net/#1}{\nolinkurl{http://ascl.net/#1}}}
\providecommand{\doarXiv}[1]{\href{https://arxiv.org/abs/#1}{\nolinkurl{https://arxiv.org/abs/#1}}}

\bibitem[{{Bellavita} {et~al.}(2022){Bellavita}, {Garc{\'\i}a}, {M{\'e}ndez}, \& {Karpouzas}}]{2022MNRAS.515.2099B}
{Bellavita}, C., {Garc{\'\i}a}, F., {M{\'e}ndez}, M., \& {Karpouzas}, K. 2022, \mnras, 515, 2099, \dodoi{10.1093/mnras/stac1922}

\bibitem[{{Belloni} \& {Hasinger}(1990)}]{1990A&A...230..103B}
{Belloni}, T., \& {Hasinger}, G. 1990, \aap, 230, 103

\bibitem[{{Belloni} {et~al.}(2020){Belloni}, {Zhang}, {Kylafis}, {Reig}, \& {Altamirano}}]{2020MNRAS.496.4366B}
{Belloni}, T.~M., {Zhang}, L., {Kylafis}, N.~D., {Reig}, P., \& {Altamirano}, D. 2020, \mnras, 496, 4366, \dodoi{10.1093/mnras/staa1843}

\bibitem[{{Blandford} {et~al.}(2019){Blandford}, {Meier}, \& {Readhead}}]{2019ARA&A..57..467B}
{Blandford}, R., {Meier}, D., \& {Readhead}, A. 2019, \araa, 57, 467, \dodoi{10.1146/annurev-astro-081817-051948}

\bibitem[{{Boller} {et~al.}(2003){Boller}, {Voges}, {Dennefeld}, {Lehmann}, {Predehl}, {Burwitz}, {Perlman}, {Gallo}, {Papadakis}, \& {Anderson}}]{2003A&A...397..557B}
{Boller}, T., {Voges}, W., {Dennefeld}, M., {et~al.} 2003, \aap, 397, 557, \dodoi{10.1051/0004-6361:20021520}

\bibitem[{{Buisson} {et~al.}(2019){Buisson}, {Fabian}, {Barret}, {F{\"u}rst}, {Gandhi}, {Garc{\'\i}a}, {Kara}, {Madsen}, {Miller}, {Parker}, {Shaw}, {Tomsick}, \& {Walton}}]{2019MNRAS.490.1350B}
{Buisson}, D.~J.~K., {Fabian}, A.~C., {Barret}, D., {et~al.} 2019, \mnras, 490, 1350, \dodoi{10.1093/mnras/stz2681}

\bibitem[{{Cabanac} {et~al.}(2010){Cabanac}, {Henri}, {Petrucci}, {Malzac}, {Ferreira}, \& {Belloni}}]{2010MNRAS.404..738C}
{Cabanac}, C., {Henri}, G., {Petrucci}, P.-O., {et~al.} 2010, \mnras, 404, 738, \dodoi{10.1111/j.1365-2966.2010.16340.x}

\bibitem[{{Dai} {et~al.}(2024){Dai}, {Kong}, {Ji}, {Zhou}, {Shui}, {Wang}, {Zhang}, \& {Santangelo}}]{2024A&A...692A.117D}
{Dai}, X., {Kong}, L., {Ji}, L., {et~al.} 2024, \aap, 692, A117, \dodoi{10.1051/0004-6361/202452132}

\bibitem[{{De Marco} {et~al.}(2013){De Marco}, {Ponti}, {Cappi}, {Dadina}, {Uttley}, {Cackett}, {Fabian}, \& {Miniutti}}]{2013MNRAS.431.2441D}
{De Marco}, B., {Ponti}, G., {Cappi}, M., {et~al.} 2013, \mnras, 431, 2441, \dodoi{10.1093/mnras/stt339}

\bibitem[{{De Marco} {et~al.}(2017){De Marco}, {Ponti}, {Petrucci}, {Clavel}, {Corbel}, {Belmont}, {Chakravorty}, {Coriat}, {Drappeau}, {Ferreira}, {Henri}, {Malzac}, {Rodriguez}, {Tomsick}, {Ursini}, \& {Zdziarski}}]{2017MNRAS.471.1475D}
{De Marco}, B., {Ponti}, G., {Petrucci}, P.~O., {et~al.} 2017, \mnras, 471, 1475, \dodoi{10.1093/mnras/stx1649}

\bibitem[{{Dragomiretskiy} \& {Zosso}(2014)}]{2014ITSP...62..531D}
{Dragomiretskiy}, K., \& {Zosso}, D. 2014, IEEE Transactions on Signal Processing, 62, 531, \dodoi{10.1109/TSP.2013.2288675}

\bibitem[{{Edwin} \& {Roberts}(1983)}]{1983SoPh...88..179E}
{Edwin}, P.~M., \& {Roberts}, B. 1983, \solphys, 88, 179, \dodoi{10.1007/BF00196186}

\bibitem[{{Franchini} {et~al.}(2023){Franchini}, {Bonetti}, {Lupi}, {Miniutti}, {Bortolas}, {Giustini}, {Dotti}, {Sesana}, {Arcodia}, \& {Ryu}}]{2023A&A...675A.100F}
{Franchini}, A., {Bonetti}, M., {Lupi}, A., {et~al.} 2023, \aap, 675, A100, \dodoi{10.1051/0004-6361/202346565}

\bibitem[{{Ghosh} {et~al.}(2023){Ghosh}, {Laha}, {Meyer}, {Roychowdhury}, {Yang}, {Acosta-Pulido}, {Rakshit}, {Pandey}, {Gonz{\'a}lez}, {Behar}, {Gallo}, {Panessa}, {Bianchi}, {La Franca}, {Scepi}, {Begelman}, {Longinotti}, {Lusso}, {Oates}, {Nicholl}, {Cenko}, {O'Connor}, {Hammerstein}, {Jose}, {Gab{\'a}nyi}, {Ricci}, \& {Chattopadhyay}}]{2023ApJ...955....3G}
{Ghosh}, R., {Laha}, S., {Meyer}, E., {et~al.} 2023, \apj, 955, 3, \dodoi{10.3847/1538-4357/aced92}

\bibitem[{{Gierli{\'n}ski} {et~al.}(2008){Gierli{\'n}ski}, {Middleton}, {Ward}, \& {Done}}]{2008Natur.455..369G}
{Gierli{\'n}ski}, M., {Middleton}, M., {Ward}, M., \& {Done}, C. 2008, \nat, 455, 369, \dodoi{10.1038/nature07277}

\bibitem[{{Giustini} {et~al.}(2020){Giustini}, {Miniutti}, \& {Saxton}}]{2020A&A...636L...2G}
{Giustini}, M., {Miniutti}, G., \& {Saxton}, R.~D. 2020, \aap, 636, L2, \dodoi{10.1051/0004-6361/202037610}

\bibitem[{{Hu} {et~al.}(2014){Hu}, {Chou}, {Yang}, \& {Su}}]{2014ApJ...788...31H}
{Hu}, C.-P., {Chou}, Y., {Yang}, T.-C., \& {Su}, Y.-H. 2014, \apj, 788, 31, \dodoi{10.1088/0004-637X/788/1/31}

\bibitem[{{Huang} \& {Wu}(2008)}]{2008RvGeo..46.2006H}
{Huang}, N.~E., \& {Wu}, Z. 2008, Reviews of Geophysics, 46, RG2006, \dodoi{10.1029/2007RG000228}

\bibitem[{{Huang} {et~al.}(1998){Huang}, {Shen}, {Long}, {Wu}, {Shih}, {Zheng}, {Yen}, {Tung}, \& {Liu}}]{1998RSPSA.454..903H}
{Huang}, N.~E., {Shen}, Z., {Long}, S.~R., {et~al.} 1998, Proceedings of the Royal Society of London Series A, 454, 903, \dodoi{10.1098/rspa.1998.0193}

\bibitem[{{Jin} {et~al.}(2021){Jin}, {Done}, \& {Ward}}]{2021MNRAS.500.2475J}
{Jin}, C., {Done}, C., \& {Ward}, M. 2021, \mnras, 500, 2475, \dodoi{10.1093/mnras/staa3386}

\bibitem[{{Kalberla} {et~al.}(2005){Kalberla}, {Burton}, {Hartmann}, {Arnal}, {Bajaja}, {Morras}, \& {P{\"o}ppel}}]{2005A&A...440..775K}
{Kalberla}, P.~M.~W., {Burton}, W.~B., {Hartmann}, D., {et~al.} 2005, \aap, 440, 775, \dodoi{10.1051/0004-6361:20041864}

\bibitem[{{Kara} {et~al.}(2016){Kara}, {Alston}, {Fabian}, {Cackett}, {Uttley}, {Reynolds}, \& {Zoghbi}}]{2016MNRAS.462..511K}
{Kara}, E., {Alston}, W.~N., {Fabian}, A.~C., {et~al.} 2016, \mnras, 462, 511, \dodoi{10.1093/mnras/stw1695}

\bibitem[{{Kara} \& {Garc{\'\i}a}(2025)}]{2025ARA&A..63..379K}
{Kara}, E., \& {Garc{\'\i}a}, J. 2025, \araa, 63, 379, \dodoi{10.1146/annurev-astro-071221-052844}

\bibitem[{{Kara} {et~al.}(2019){Kara}, {Steiner}, {Fabian}, {Cackett}, {Uttley}, {Remillard}, {Gendreau}, {Arzoumanian}, {Altamirano}, {Eikenberry}, {Enoto}, {Homan}, {Neilsen}, \& {Stevens}}]{2019Natur.565..198K}
{Kara}, E., {Steiner}, J.~F., {Fabian}, A.~C., {et~al.} 2019, \nat, 565, 198, \dodoi{10.1038/s41586-018-0803-x}

\bibitem[{{Kato}(1990)}]{1990PASJ...42...99K}
{Kato}, S. 1990, \pasj, 42, 99, \dodoi{10.1093/pasj/42.1.99}

\bibitem[{{Kazanas} {et~al.}(1997){Kazanas}, {Hua}, \& {Titarchuk}}]{1997ApJ...480..735K}
{Kazanas}, D., {Hua}, X.-M., \& {Titarchuk}, L. 1997, \apj, 480, 735, \dodoi{10.1086/303991}

\bibitem[{{Laha} {et~al.}(2022){Laha}, {Meyer}, {Roychowdhury}, {Becerra Gonzalez}, {Acosta-Pulido}, {Thapa}, {Ghosh}, {Behar}, {Gallo}, {Kriss}, {Panessa}, {Bianchi}, {La Franca}, {Scepi}, {Begelman}, {Longinotti}, {Lusso}, {Oates}, {Nicholl}, \& {Cenko}}]{2022ApJ...931....5L}
{Laha}, S., {Meyer}, E., {Roychowdhury}, A., {et~al.} 2022, \apj, 931, 5, \dodoi{10.3847/1538-4357/ac63aa}

\bibitem[{{Laha} {et~al.}(2025){Laha}, {Meyer}, {Sadaula}, {Ghosh}, {Sengupta}, {Masterson}, {Shuvo}, {Guainazzi}, {Ricci}, {Begelman}, {Philippov}, {Mbarek}, {Hankla}, {Kara}, {Panessa}, {Behar}, {Zhang}, {Pacucci}, {Pal}, {Ricci}, {Villani}, {Bisogni}, {La Franca}, {Bianchi}, {Bruni}, {Oates}, {Hahn}, {Nicholl}, {Cenko}, {Chattopadhyay}, {Becerra Gonz{\'a}lez}, {Acosta{\textendash}Pulido}, {Rakshit}, {Svoboda}, {Gallo}, {Ingram}, \& {Kakkad}}]{2025ApJ...981..125L}
{Laha}, S., {Meyer}, E.~T., {Sadaula}, D.~R., {et~al.} 2025, \apj, 981, 125, \dodoi{10.3847/1538-4357/adaea0}

\bibitem[{{Li} {et~al.}(2024){Li}, {Ho}, {Ricci}, \& {Trakhtenbrot}}]{2024ApJ...975...50L}
{Li}, R., {Ho}, L.~C., {Ricci}, C., \& {Trakhtenbrot}, B. 2024, \apj, 975, 50, \dodoi{10.3847/1538-4357/ad77a5}

\bibitem[{{Li} {et~al.}(2022){Li}, {Ho}, {Ricci}, {Trakhtenbrot}, {Arcavi}, {Kara}, \& {Hiramatsu}}]{2022ApJ...933...70L}
{Li}, R., {Ho}, L.~C., {Ricci}, C., {et~al.} 2022, \apj, 933, 70, \dodoi{10.3847/1538-4357/ac714a}

\bibitem[{{Lin} {et~al.}(2013){Lin}, {Irwin}, {Godet}, {Webb}, \& {Barret}}]{2013ApJ...776L..10L}
{Lin}, D., {Irwin}, J.~A., {Godet}, O., {Webb}, N.~A., \& {Barret}, D. 2013, \apjl, 776, L10, \dodoi{10.1088/2041-8205/776/1/L10}

\bibitem[{{Linial} \& {Metzger}(2023)}]{2023ApJ...957...34L}
{Linial}, I., \& {Metzger}, B.~D. 2023, \apj, 957, 34, \dodoi{10.3847/1538-4357/acf65b}

\bibitem[{{Ma} {et~al.}(2023){Ma}, {M{\'e}ndez}, {Garc{\'\i}a}, {Sai}, {Zhang}, \& {Zhang}}]{2023MNRAS.525..854M}
{Ma}, R., {M{\'e}ndez}, M., {Garc{\'\i}a}, F., {et~al.} 2023, \mnras, 525, 854, \dodoi{10.1093/mnras/stad2284}

\bibitem[{{Masterson} {et~al.}(2022){Masterson}, {Kara}, {Ricci}, {Garc{\'\i}a}, {Fabian}, {Pinto}, {Kosec}, {Remillard}, {Loewenstein}, {Trakhtenbrot}, \& {Arcavi}}]{2022ApJ...934...35M}
{Masterson}, M., {Kara}, E., {Ricci}, C., {et~al.} 2022, \apj, 934, 35, \dodoi{10.3847/1538-4357/ac76c0}

\bibitem[{{Masterson} {et~al.}(2025){Masterson}, {Kara}, {Panagiotou}, {Alston}, {Chakraborty}, {Burdge}, {Ricci}, {Laha}, {Arcavi}, {Arcodia}, {Cenko}, {Fabian}, {Garc{\'\i}a}, {Giustini}, {Ingram}, {Kosec}, {Loewenstein}, {Meyer}, {Miniutti}, {Pinto}, {Remillard}, {Sadaula}, {Shuvo}, {Trakhtenbrot}, \& {Wang}}]{2025Natur.638..370M}
{Masterson}, M., {Kara}, E., {Panagiotou}, C., {et~al.} 2025, \nat, 638, 370, \dodoi{10.1038/s41586-024-08385-x}

\bibitem[{{M{\'e}ndez} {et~al.}(2013){M{\'e}ndez}, {Altamirano}, {Belloni}, \& {Sanna}}]{2013MNRAS.435.2132M}
{M{\'e}ndez}, M., {Altamirano}, D., {Belloni}, T., \& {Sanna}, A. 2013, \mnras, 435, 2132, \dodoi{10.1093/mnras/stt1431}

\bibitem[{{Meyer} {et~al.}(2025){Meyer}, {Laha}, {Shuvo}, {Roychowdhury}, {Green}, {Rhodes}, {Hankla}, {Philippov}, {Mbarek}, {laor}, {Begelman}, {Sadaula}, {Ghosh}, {Bruni}, {Panessa}, {Guainazzi}, {Behar}, {Masterson}, {Zhang}, {Yang}, {Gurwell}, {Keating}, {Williams-Baldwin}, {Bray}, {Bempong-Manful}, {Wrigley}, {Bianchi}, {Ricci}, {La Franca}, {Kara}, {Georganopoulos}, {Oates}, {Nicholl}, {Pal}, \& {Cenko}}]{2025ApJ...979L...2M}
{Meyer}, E.~T., {Laha}, S., {Shuvo}, O.~I., {et~al.} 2025, \apjl, 979, L2, \dodoi{10.3847/2041-8213/ad8651}

\bibitem[{{Miniutti} {et~al.}(2019){Miniutti}, {Saxton}, {Giustini}, {Alexander}, {Fender}, {Heywood}, {Monageng}, {Coriat}, {Tzioumis}, {Read}, {Knigge}, {Gandhi}, {Pretorius}, \& {Ag{\'\i}s-Gonz{\'a}lez}}]{2019Natur.573..381M}
{Miniutti}, G., {Saxton}, R.~D., {Giustini}, M., {et~al.} 2019, \nat, 573, 381, \dodoi{10.1038/s41586-019-1556-x}

\bibitem[{{Miyamoto} {et~al.}(1988){Miyamoto}, {Kitamoto}, {Mitsuda}, \& {Dotani}}]{1988Natur.336..450M}
{Miyamoto}, S., {Kitamoto}, S., {Mitsuda}, K., \& {Dotani}, T. 1988, \nat, 336, 450, \dodoi{10.1038/336450a0}

\bibitem[{Nazari \& Sakhaei(2018)}]{2018IJBHI...22.1059N}
Nazari, M., \& Sakhaei, S.~M. 2018, IEEE Journal of Biomedical and Health Informatics, 22, 1059

\bibitem[{{Nowak} {et~al.}(1999){Nowak}, {Vaughan}, {Wilms}, {Dove}, \& {Begelman}}]{1999ApJ...510..874N}
{Nowak}, M.~A., {Vaughan}, B.~A., {Wilms}, J., {Dove}, J.~B., \& {Begelman}, M.~C. 1999, \apj, 510, 874, \dodoi{10.1086/306610}

\bibitem[{{Panessa} {et~al.}(2019){Panessa}, {Baldi}, {Laor}, {Padovani}, {Behar}, \& {McHardy}}]{2019NatAs...3..387P}
{Panessa}, F., {Baldi}, R.~D., {Laor}, A., {et~al.} 2019, Nature Astronomy, 3, 387, \dodoi{10.1038/s41550-019-0765-4}

\bibitem[{{Pasham} {et~al.}(2019){Pasham}, {Remillard}, {Fragile}, {Franchini}, {Stone}, {Lodato}, {Homan}, {Chakrabarty}, {Baganoff}, {Steiner}, {Coughlin}, \& {Pasham}}]{2019Sci...363..531P}
{Pasham}, D.~R., {Remillard}, R.~A., {Fragile}, P.~C., {et~al.} 2019, Science, 363, 531, \dodoi{10.1126/science.aar7480}

\bibitem[{{Peirano} {et~al.}(2023){Peirano}, {M{\'e}ndez}, {Garc{\'\i}a}, \& {Belloni}}]{2023MNRAS.519.1336P}
{Peirano}, V., {M{\'e}ndez}, M., {Garc{\'\i}a}, F., \& {Belloni}, T. 2023, \mnras, 519, 1336, \dodoi{10.1093/mnras/stac3553}

\bibitem[{{Popham} {et~al.}(1999){Popham}, {Woosley}, \& {Fryer}}]{1999ApJ...518..356P}
{Popham}, R., {Woosley}, S.~E., \& {Fryer}, C. 1999, \apj, 518, 356, \dodoi{10.1086/307259}

\bibitem[{{Rees}(1988)}]{1988Natur.333..523R}
{Rees}, M.~J. 1988, \nat, 333, 523, \dodoi{10.1038/333523a0}

\bibitem[{{Ricci} {et~al.}(2020){Ricci}, {Kara}, {Loewenstein}, {Trakhtenbrot}, {Arcavi}, {Remillard}, {Fabian}, {Gendreau}, {Arzoumanian}, {Li}, {Ho}, {MacLeod}, {Cackett}, {Altamirano}, {Gandhi}, {Kosec}, {Pasham}, {Steiner}, \& {Chan}}]{2020ApJ...898L...1R}
{Ricci}, C., {Kara}, E., {Loewenstein}, M., {et~al.} 2020, \apjl, 898, L1, \dodoi{10.3847/2041-8213/ab91a1}

\bibitem[{{Salpeter}(1964)}]{1964ApJ...140..796S}
{Salpeter}, E.~E. 1964, \apj, 140, 796, \dodoi{10.1086/147973}

\bibitem[{{Shakura} \& {Sunyaev}(1973)}]{1973A&A....24..337S}
{Shakura}, N.~I., \& {Sunyaev}, R.~A. 1973, \aap, 24, 337

\bibitem[{{Shui} {et~al.}(2023){Shui}, {Zhang}, {Zhang}, {Chen}, {Kong}, {Wang}, {Peng}, {Ji}, {Santangelo}, {Yin}, {Qu}, {Tao}, {Ge}, {Huang}, {Zhang}, {Liu}, {Zhang}, {Yu}, {Chang}, {Li}, {Ye}, {Li}, {Yu}, \& {Yan}}]{2023ApJ...957...84S}
{Shui}, Q.~C., {Zhang}, S., {Zhang}, S.~N., {et~al.} 2023, \apj, 957, 84, \dodoi{10.3847/1538-4357/acfc42}

\bibitem[{{Shui} {et~al.}(2024{\natexlab{a}}){Shui}, {Zhang}, {Zhang}, {Chen}, {Kong}, {Peng}, {Ji}, {Wang}, {Chang}, {Yu}, {Yin}, {Qu}, {Tao}, {Ge}, {Ma}, {Zhang}, {Yu}, \& {Li}}]{2024ApJ...965L...7S}
---. 2024{\natexlab{a}}, \apjl, 965, L7, \dodoi{10.3847/2041-8213/ad374d}

\bibitem[{{Shui} {et~al.}(2024{\natexlab{b}}){Shui}, {Zhang}, {Peng}, {Zhang}, {Chen}, {Ji}, {Kong}, {Feng}, {Yu}, {Wang}, {Chang}, {Yin}, {Qu}, {Tao}, {Ge}, {Zhang}, \& {Li}}]{2024ApJ...973...59S}
{Shui}, Q.-C., {Zhang}, S., {Peng}, J.-Q., {et~al.} 2024{\natexlab{b}}, \apj, 973, 59, \dodoi{10.3847/1538-4357/ad676a}

\bibitem[{{Tagger} \& {Pellat}(1999)}]{1999A&A...349.1003T}
{Tagger}, M., \& {Pellat}, R. 1999, \aap, 349, 1003, \dodoi{10.48550/arXiv.astro-ph/9907267}

\bibitem[{{Trakhtenbrot} {et~al.}(2019){Trakhtenbrot}, {Arcavi}, {MacLeod}, {Ricci}, {Kara}, {Graham}, {Stern}, {Harrison}, {Burke}, {Hiramatsu}, {Hosseinzadeh}, {Howell}, {Smartt}, {Rest}, {Prieto}, {Shappee}, {Holoien}, {Bersier}, {Filippenko}, {Brink}, {Zheng}, {Li}, {Remillard}, \& {Loewenstein}}]{2019ApJ...883...94T}
{Trakhtenbrot}, B., {Arcavi}, I., {MacLeod}, C.~L., {et~al.} 2019, \apj, 883, 94, \dodoi{10.3847/1538-4357/ab39e4}

\bibitem[{{Uttley} {et~al.}(2014){Uttley}, {Cackett}, {Fabian}, {Kara}, \& {Wilkins}}]{2014A&ARv..22...72U}
{Uttley}, P., {Cackett}, E.~M., {Fabian}, A.~C., {Kara}, E., \& {Wilkins}, D.~R. 2014, \aapr, 22, 72, \dodoi{10.1007/s00159-014-0072-0}

\bibitem[{{Vaughan}(2010)}]{2010MNRAS.402..307V}
{Vaughan}, S. 2010, \mnras, 402, 307, \dodoi{10.1111/j.1365-2966.2009.15868.x}

\bibitem[{{Wagoner}(1999)}]{1999PhR...311..259W}
{Wagoner}, R.~V. 1999, \physrep, 311, 259, \dodoi{10.1016/S0370-1573(98)00104-5}

\bibitem[{{Wang} {et~al.}(2020){Wang}, {Kara}, {Steiner}, {Garc{\'\i}a}, {Homan}, {Neilsen}, {Marcel}, {Ludlam}, {Tombesi}, {Cackett}, \& {Remillard}}]{2020ApJ...899...44W}
{Wang}, J., {Kara}, E., {Steiner}, J.~F., {et~al.} 2020, \apj, 899, 44, \dodoi{10.3847/1538-4357/ab9ec3}

\bibitem[{{Zhang} {et~al.}(2019){Zhang}, {Dov{\v{c}}iak}, \& {Bursa}}]{2019ApJ...875..148Z}
{Zhang}, W., {Dov{\v{c}}iak}, M., \& {Bursa}, M. 2019, \apj, 875, 148, \dodoi{10.3847/1538-4357/ab1261}

\bibitem[{{Zhao} {et~al.}(2024){Zhao}, {Tao}, {Li}, {Zhang}, {Feng}, {Ge}, {Ji}, {Wang}, {Huang}, {Ma}, {Zhang}, {Qu}, {Xu}, {Zhang}, {Yin}, {Shui}, {Ma}, {Zhao}, {Li}, {Yang}, {Liu}, \& {Yu}}]{2024ApJ...961L..42Z}
{Zhao}, Q.-C., {Tao}, L., {Li}, H.-C., {et~al.} 2024, \apjl, 961, L42, \dodoi{10.3847/2041-8213/ad1e6c}

\bibitem[{{Zhou} {et~al.}(2015){Zhou}, {Yuan}, {Pan}, \& {Liu}}]{2015ApJ...798L...5Z}
{Zhou}, X.-L., {Yuan}, W., {Pan}, H.-W., \& {Liu}, Z. 2015, \apjl, 798, L5, \dodoi{10.1088/2041-8205/798/1/L5}

\end{thebibliography}
\bibliographystyle{aasjournal}



\end{document}